\documentclass[onecolumn]{aastex631}
\usepackage{mathtools}
\usepackage{amsmath}
\usepackage{soul}
\usepackage{lipsum}
\usepackage{gensymb}
\usepackage{rotating}
\usepackage{multirow}

\newcommand\kms{$\mathrm{km\ s^{-1}}$}
\newcommand\Rsun{$\mathrm{R_{\odot}}$}
\newcommand\Rss{$\mathrm{R_{ss}}$}
\newcommand\car{$\mathrm{C^{6+}/C^{4+}}$}

\newcommand\oxy{$\mathrm{O^{7+}/O^{6+}}$}
\newcommand\feo{$\mathrm{Fe/O}$}
\newcommand\Ahe{$\mathrm{A_{He}}$}
\newcommand\vap{$\mathrm{\Delta v_{\alpha p}}$}
\newcommand\MA{$\mathrm{M_A}$}

\newcommand\Bo{$\mathrm{B_0}$}
\newcommand\sigmac{$\mathrm{\sigma_C}$}
\newcommand\sigmar{$\mathrm{\sigma_R}$}
\newcommand\fss{$\mathrm{f_{ss}}$}

\newcommand\vsw{$\mathrm{v_{sw}}$}
\newcommand\alf{Alfv\'en}
\newcommand\alfic{Alfv\'enic}
\newcommand\alfty{Alfv\'enicity}

\newcommand\elas{Els\"{a}sser}

\newcommand\saswnum{10} % number of in depth modeled SASW streams
 
\newcommand\fswnum{5} 

\shorttitle{Slow Alfv\'enic Solar Wind}
\shortauthors{Ervin et al.}

\begin{document}

\title{Characteristics and Source Regions of Slow Alfv\'enic Solar Wind Observed by Parker Solar Probe}

\correspondingauthor{Tamar Ervin}
\email{tamarervin@berkeley.edu}

\author[0000-0002-8475-8606]{Tamar Ervin}
\affiliation{Department of Physics, University of California, Berkeley, Berkeley, CA 94720-7300, USA}
\affiliation{Space Sciences Laboratory, University of California, Berkeley, CA 94720-7450, USA}

\author[0009-0008-6049-255X]{Kai Jaffarove}
\affiliation{Department of Physics, University of California, Berkeley, Berkeley, CA 94720-7300, USA}
\affiliation{Space Sciences Laboratory, University of California, Berkeley, CA 94720-7450, USA}

\author[0000-0002-6145-436X]{Samuel T. Badman}
\affiliation{Center for Astrophysics $\vert$ Harvard \& Smithsonian, 60 Garden Street, Cambridge, MA 02138, USA}

\author[0000-0002-9954-4707]{Jia Huang}
\affiliation{Space Sciences Laboratory, University of California, Berkeley, CA 94720-7450, USA}

\author[0000-0002-8748-2123]{Yeimy J. Rivera} 
\affiliation{Center for Astrophysics $\vert$ Harvard \& Smithsonian, 60 Garden Street, Cambridge, MA 02138, USA}

\author[0000-0002-1989-3596]{Stuart D. Bale}
\affiliation{Department of Physics, University of California, Berkeley, Berkeley, CA 94720-7300, USA}
\affiliation{Space Sciences Laboratory, University of California, Berkeley, CA 94720-7450, USA}

%%%%%%% ------------------------------------------------------------------------------------------------------------------------------ %%%%%%%
%%%%%%% -------------------------------------------------------- ABSTRACT -------------------------------------------------------- %%%%%%%
%%%%%%% ------------------------------------------------------------------------------------------------------------------------------ %%%%%%%
\begin{abstract}
Using a classification scheme for solar wind type based on the heliocentric distance of the observation, we look at near perihelion observations from Parker Solar Probe Encounters Four to Fourteen to study the sources of the slow {\alfic} solar wind (SASW). Through Potential Field Source Surface (PFSS) modeling and ballistic mapping, we connect streams to their solar source and find that a primary population of SASW comes from low magnetic field strength regions (low-{\Bo}), likely small coronal holes (CHs) and their over-expanded boundaries, while a second population of high field strength (high-{\Bo}) seems to emerge from non-CH structures potentially through interchange reconnection with nearby open field lines. This low-{\Bo} SASW shows larger expansion than the fast solar wind (FSW) but similar mass flux, potentially indicating additional heating below the critical point, and emergence from a cooler structure, which could lead to slower wind emerging from CH-like structures. We show that this low-{\Bo} SASW shows stronger preferential acceleration of alpha particles (similar to the FSW) than the high-{\Bo} SASW, and that this is a velocity dependent phenomenon as found in previous studies. To have additional confidence in our mapping results, we quantify the error on both the PFSS model and ballistic mapping and discuss how additional multi-point observations of plasma parameters and composition would allow us to better constrain our models and connect the solar wind to its source. 
\end{abstract}

% Additionally, this CH-like SASW shows higher mass flux and expansion that FSW from CHs indicating additional heating below the critical point leading to slower wind emerging from CH-like structures. 
% Key Points:
% \begin{itemize}
%    \item some of the SASW comes from CH/boundaries 
%    \item some of it comes from non-CH structures, this stuff has predominately {\MA} $<$ 1 and could be released from closed loop structures through interchange reconnection
%    \item {\Ahe} look similar between SASW from different sources but fundamentally different than FSW
%    \item PFSS modeling methods make it difficult to identify wind emerging from non-CH structures (trace along open field lines) however with composition metrics we can potentially make some guesses
% \end{itemize}

%%%%%%% ------------------------------------------------------------------------------------------------------------------------------ %%%%%%%
%%%%%%% -------------------------------------------------------- INTRODUCTION -------------------------------------------------------- %%%%%%%
%%%%%%% ------------------------------------------------------------------------------------------------------------------------------ %%%%%%%
\section{Introduction} \label{sec: intro}

% intro to types of solar wind
The source regions of the solar wind have been a topic of interest to the space physics community for decades, and while the source of the fast solar wind (FSW) is well known to be coronal holes (CH) \citep[and others]{McComas-1998, McComas-2008}, the source of the slow solar wind is still unresolved. The solar wind has traditionally been categorized by wind speed at 1 AU: fast (speeds above 500 {\kms} at 1 AU) and slow (speeds below 500 {\kms}), however more recent works have shown that the slow solar wind can be further categorized into slow {\alfic} solar wind (SASW) and classical slow solar wind (SSW) based on the {\alfty} of the plasma \citep[and others]{DAmicis-2015, Stansby-2018, Stansby-2020alf, Ervin-2024CH}. This SASW shows similar properties to the fast solar wind specifically in terms of {\alfty} \citep{DAmicis-2015, Perrone-2020, DAmicis-2021}, alpha particle abundances \citep{Ohmi-2004, Ervin-2024CH}, differential velocities \citep{Stansby-2020comp}, and composition \citep{Stakhiv-2015, DAmicis-2018, Ervin-2024CH} meaning that this wind likely has a similar source to the FSW \citep[and others]{DAmicis-2015, DAmicis-2021}.

% intro to source regions
Since the FSW emerges from open field lines streaming from within CHs, some of the SASW likely originates from CHs and/or their boundaries \citep[and others]{Wang-1990}. Expansion within these CH regions can play a role in the speed of the emergent wind and perhaps lead to slower wind emerging from CH regions. Specifically, heating above the critical point (where the wind becomes supersonic) increases wind speeds \citep{Leer-1980}. Heating concentrated near the coronal base leads to a larger mass flux and lower energy available per particle which decreases the acceleration leading to slower wind speeds \citep{Wang-2009}. Therefore, rapidly diverging fields with higher critical points and less energy deposited near the sonic point can produce slower speed winds \citep{Wang-2012, Stansby-2020alf}. \citet{Chen-2001, Hu-2003} study the acceleration of the solar wind using a 2D {\alf}-wave driven model, finding that the geometry (expansion) of the flow tube is important in the formation of slow wind.

% magnetic expansion
The magnetic expansion factor ($f_{s}$) is a measure of the divergence of the magnetic field \citep{Wang-1990, Wang-1997} and there is an empirically established anti-correlation between the expansion factor and solar wind speed \citep{Wang-1990}. Small CHs and their boundaries have been shown to have higher magnetic expansion \citep{Wallace-2020, Ervin-2024CH} and could therefore be a source of slower wind due to higher critical points leading to heating below the transition to supersonic wind speeds \citep{Nolte-1976, Garton-2018}. It has been shown that the expansion factor does not always follow this correlation with wind speed, but that the dynamic boundary between open and closed magnetic field lines could be responsible for this correlation or deviation from it \citep{Riley-2015, Wallace-2020}. Recent work by \citet{Yardley-2024} combined spectroscopic observations with in situ measurements of plasma parameters and composition to find the emergence of wind streams from an AR-CH complex with an over-expanded magnetic field. They claim that coronal plasma trapped within closed loops can be released along open fan loops via interchange reconnection.

% composition
In addition to the correlation between expansion factor and wind type, different compositional characteristics have been shown to be associated with different wind speeds and a potential tracer of the source region of the wind. Due to the \textit{freeze-in} process, composition ratios stay constant through the radial evolution of the solar wind and are thus a good metric to evaluate the source of the wind when compared with known properties of various source regions \citep{Buergi-1986, Chen-2003, Landi-2012b}. The processes that lead to compositional variations occur in the solar atmosphere itself, meaning that the amount of time spent in the corona will lead to differences in compositional metrics, and plasma must stay in the corona long enough (2-3 days) to experience these fractionation effects \citep{Laming-2015}. Therefore, plasma quickly escaping the corona on open field lines (such as from CH structures) show more \lq{}photospheric\rq{} like compositions in comparison to plasma escaping from closed field lines or loop-like structures \citep{Feldman-1997, Laming-2017}. These plasma parcels would show an enhancement in low first ionization potential (FIP) elements (high {\feo} ratio) and enhanced ion ratios arising from the hotter source region \citep{Geiss-1995, vonSteiger-2000}. 

% alpha particles
The solar wind typically shows alpha particle (Helium; He) abundance ratios ({\Ahe}) that are depleted compared with photospheric values ($\sim$8.4\%, \citet{Asplund-2021}) and are highly variable. Helium rich periods are often associated with solar eruptions \citep{Song-2022, Feldman-2005, Johnson-2024}, streamer belt outflows, and other active region structures \citep{Kasper-2016, Alterman-2018, Alterman-2019}. There is some modulation of the abundance with the solar cycle which could be due to variations in the structure of the solar wind source regions throughout the solar cycle \citep{Norton-2023}. Differences in {\Ahe} can also be attributed to gravitational settling, another process that shows solar cycle dependence \citep{Rivera-2022}. Therefore, even though Helium is the highest FIP element, the ubiquitously observed depletion of Helium within the ambient solar wind indicates that the process(es) leading to its enhancement or depletion are likely different from the FIP effect as described above \citep{Johnson-2024} and thus there is no full picture to describe what we would expect in terms of alpha abundance related to source region. Typically, {\Ahe} are expected to be enhanced in plasma streaming from open field lines or CH-like structures, however this does not always hold true \citep{Ervin-2024CH}. It is important to note that due to alpha-proton differential streaming, {\Ahe} does not necessarily remain constant throughout solar wind evolution \citep{Zhang-2024}.

% These plasma parcels would show an enhancement in low first ionization potential (FIP) elements and ionization rates leading to higher ratios of measured ratios such as {\oxy}, {\car}, and {\feo} and decreased alpha particle abundance ratios as Helium is a high FIP element compared to the low FIP hydrogen \citep{Geiss-1995, vonSteiger-2000}. 
% \red{As Parker Solar Probe does not contain a thermal ion composition instrument, in this study we will focus on the alpha particle abundance ratio, and point out that including ion composition instruments on future inner heliospheric missions is vital to improve our understanding of solar wind source regions and the processes that lead to compositional variations both within the solar atmosphere and through solar wind propagation through the heliosphere.}

% paper overview
In this study, we use Parker Solar Probe (Parker) observations within 40~{\Rsun} from Encounters 4 to 14 to study the in situ characteristics and source regions of the SASW in comparison with the FSW, as the source region and characteristics of the FSW are well understood. We will not make any definitive claims regarding the general source of the SASW, but rather seek to highlight that they likely emerge from a variety of sources -- some similar to FSW, some different. In Sections~\ref{sec: data} and~\ref{sec: methods}, we outline our data selection and wind stream identification methods along with modeling methods we use to identify the source regions of the SASW streams of interest (additional details can be found in Appendix~\ref{appendix: velocity}). Section~\ref{sec: disc} discusses the in situ and modeling parameters associated with the SASW and identified source regions. We conclude in Section~\ref{sec: conclusion} with an overview of our results, the limitations of this study, and future work necessary to fully understand the source regions of the slow {\alfic} solar wind. In Appendix~\ref{appendix: streams} we show overviews of modeling results for specific streams, and in Appendix~\ref{appendix: validation} discuss uncertainty quantification on our modeling methods.

%%%%%%% ------------------------------------------------------------------------------------------------------------------------------ %%%%%%%
%%%%%%% -------------------------------------------------------- DATA -------------------------------------------------------- %%%%%%%
\section{Data Overview} \label{sec: data}
% Parker
This study uses observations within 40~{\Rsun} during Encounters 4 to 14 (E4 - 14) of Parker Solar Probe (Parker) spanning January 2020 to January 2023, providing a good overview of the solar wind at the inner heliosphere during solar minimum and into the rising phase of the solar cycle. We use near perihelion magnetic field measurements from the Electromagnetic Fields Investigation (FIELDS; \citet{Bale-2016}) fluxgate magnetometer and partial plasma moments of alphas and protons from the ion Solar Probe ANalyzer (SPAN-I; \citet{Livi-2022}) aboard the \lq{}Solar Wind Electrons, Alphas, and Protons\rq{} (SWEAP; \citet{Kasper-2016}) suite. SPAN-I measurements are available starting in Encounter 4 which is the reason observations from E1 to 3 are not used such that we can maintain consistency throughout our study, and use the native quality flags to mitigate field-of-view (FOV) issues. An important note is that the SPAN-I absolute abundance measurements (number density) is likely an underestimate of the total density as they are computed from partial moments \citep{Livi-2022}.

% GONG, AIA
As the lower boundary photospheric magnetic field observation to our potential field source surface (PFSS) modeling (described in Section~\ref{sec: methods-modeling}), we use ADAPT-GONG magnetograms. These are high resolution, full disk photospheric magnetograms from the Global Oscillation Network Group (GONG; \citet{Harvey-1996}) with the Air Force Data Assimilative Photospheric Flux Transport (ADAPT; \citet{Worden-2000, Arge-2010, Arge-2011, Arge-2013, Hickmann-2015}) model applied. To compare footpoints estimated from PFSS method, we use EUV observations from the Atmospheric Imaging Assembly (AIA; \citet{Lemen-2012}) aboard the Solar Dynamics Observatory (SDO; \citet{Pesnell-2012}), which takes high resolution and high cadence, narrow band full disk images of the solar atmosphere. We look at SDO/AIA observations of 193~{\AA} allowing us to probe EUV emissions in the corona with a characteristic $\log{T}$ of 6.3 \citep{Lemen-2012}. 

% maybe include an example of plasma parameters

%%%%%%% ------------------------------------------------------------------------------------------------------------------------------ %%%%%%%
%%%%%%% -------------------------------------------------------- METHODS -------------------------------------------------------- %%%%%%%
%%%%%%% ------------------------------------------------------------------------------------------------------------------------------ %%%%%%%
\section{Methods} \label{sec: methods}

%%%%%%% -------------- Stream Identification -------------- %%%%%%%
\subsection{Stream Identification} \label{sec: methods-identification} % [Kai]

All usable (good field of view) E4 - E14 data within 40~{\Rsun} was compiled and analyzed based on certain parameters. The objective of the method is to classify observed wind as SSW, SASW, or FSW. We can detect all three wind types simultaneously because the designated parameters are mutually exclusive such that each observation can belong only to a specific wind category.

% describe identification of SASW, FSW, SSW streams
It is common practice to identify slow solar wind as streams with mean speeds below a threshold value (typically 400 to 500~{\kms} depending on the heliocentric distance and study) for a specified time period. In this case, due to Parker being so close to the Sun, we modified the traditional classification scheme to account for solar wind acceleration. Using observations over more than a full solar cycle (2004 to 2024) from the Wind spacecraft \citep{Lin-1995aa} at 1 AU, we looked at the fraction of the observed wind that would be considered classically fast. We then narrow this window to just the time period of the Parker observations we look at (January 2020 to January 2023) finding that $\sim 20 \%$ was above 500~{\kms} and $\sim 50 \%$ above 400~{\kms} at 1 AU (Figure~\ref{fig: wind_figure}). Appendix~\ref{appendix: velocity} shows an overview of this method and how these percentage thresholds stay consistent throughout time, such that we can apply the categorization thresholds to Parker data observed at different radial distances. 

\begin{figure} [ht] %[htb!]
\begin{center}
  \includegraphics[width=\columnwidth]{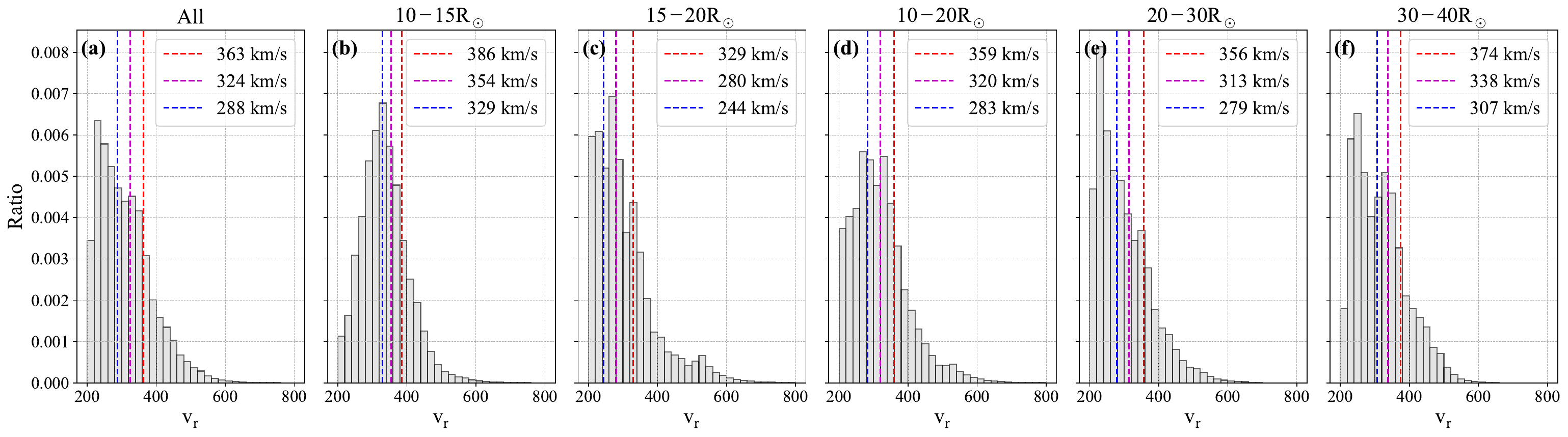}
  \caption{Distributions of solar wind speed binned by heliocentric distance. The dashed lines indicate different threshold percentages based on observations of wind speeds at 1 AU (see section~\ref{sec: methods-identification}) as applied to Parker. The red, purple, and blue lines signify the uppermost 20\%, 35\%, and 50\% of the wind respectively where 20\% is the 500~{\kms} cutoff at 1 AU, 50\% is the 400~{\kms} cutoff, and 35\% falls in between. In our data stream identification, we use the red line, or the equivalent threshold to 500~{\kms} at 1 AU from Parker to differentiate between slow and fast solar wind. The velocities associated with each dashed line are listed.
  \textit{Panel (a):} All Parker solar wind velocity data below 40~{\Rsun} for Encounters 4-14 with associated dashed lines.
   \textit{Panels (b) and (c):} Parker solar wind velocity data for radial bin (b) 10-15~{\Rsun} and (c) 15-20~{\Rsun}.
  \textit{Panel (d) - (f):} All Parker solar wind velocity data for radial bins (d): 10-20~{\Rsun}, (e): 20-30~{\Rsun} and (f): 30-40~{\Rsun} with the associated thresholds.
  }
  \label{fig: velocity_histogram}
\end{center}
\end{figure}

Binning by radial distance in bins of 10~{\Rsun} from 10 to 40~{\Rsun} gave different thresholds for the fast versus slow solar wind cutoff {based on heliocentric distance. Wind is classified as \lq{}slow\rq{} when speeds fall below the 500 {\kms} equivalent percentage threshold (20\% or the red dashed line in Figure~\ref{fig: velocity_histogram}).} Works by \citet{Dakeyo-2022, Halekas-2022, Halekas-2023, Rivera-2024} show significant solar wind acceleration within radial distances that are relevant to this paper, often increasing in speed enough to shift between traditional slow into fast wind, providing strong motivation for our speed classification scheme. It is also noteworthy that Parker solar wind solutions do not asymptote mathematically \citep{Parker-1958} and all wind solutions continue gradual acceleration throughout the inner heliosphere.

For 10-20~{\Rsun} bin, we separate into bins of 5~{\Rsun} for further investigation of the threshold speed. We note that the innermost bin (10-15~{\Rsun}) has speeds that are significantly higher than other bins. The wind sampled here is primarily at the same radial distance and the most extreme latitudes of Parker's orbit, for shorter periods of time and thus we believe that this higher threshold could be due to sampling bias. In addition, Parker only reaches these distances during the later orbits and there are about half as many observations within this radial bin as compared to the others. Therefore, we choose to use the 15-20~{\Rsun} threshold and apply it to all observations at distances within 20~{\Rsun}.

This gives thresholds of 329~{\kms}, 356~{\kms}, and 374~{\kms} for the bins associated with heliocentric distances of 10-20~{\Rsun}, 20-30~{\Rsun}, and 30-40~{\Rsun} respectively. Applying these new thresholds for each radial bin, rather than the traditional classification, attempts to account for potential slow wind acceleration into fast wind beyond the location of Parker. As a note, we removed the September 5, 2022 coronal mass ejection (CME) event from the dataset based on Table 1 in \citet{Romeo-2023}. This is one of the largest CMEs observed to date and impacted the parameters associated with the FSW category. In general there is no removal of other CME or transient events and we expect that CMEs make up a small fraction of the solar wind measurements given the time frame of the Parker observations are near solar minimum.

% Applying these new thresholds for each radial bin rather than a fixed value of threshold speed helps account for the affects of solar wind acceleration at distances beyond the location of Parker, and attempts to account for slow wind which can potentially accelerate to FSW by the time it reaches Earth.

%% FIGURE --- velocity histogram

% red line signifies the uppermost 13\% of solar wind measured from Earth with mean speeds above 500 km/s. The blue line signifies the uppermost 38\% of solar wind measured from Earth with mean speeds above 400 km/s. The purple line indicates a median point of the uppermost 25\% between the two thresholds. 

% SSW was classified with mean speeds below 400 km/s and mean {\sigmac} values below 0.7, and FSW was classified with mean speeds above 400 km/s and mean {\sigmac} values above 0.7. 

% For additional precision, we referred to data on the solar wind composition measured from Earth and found that around 13 percent of the solar wind measured at 1 AU was classified as FSW with mean speeds above 500 km/s. 
% We applied this percentage of FSW to the data measured from Parker during perihelion to account for wind that potentially hasn't fully accelerated yet. 

In addition to categorizing the wind by speed, we further categorize the slow wind into classically slow and slow {\alfic} wind based on the cross helicity ({\sigmac}). {\sigmac} is generally used as a proxy for the {\alfty} of the plasma where higher cross helicity measurements ({\sigmac} values of 1 or -1) are indicative of pure {\alf} waves propagating parallel or anti-parallel to the magnetic field. {\sigmac} is calculated via Equation~\ref{eqn: sigmac} where $E^\pm$ is the energy associated with the {\elas} variables: $\mathbf{z}^\pm = \delta \mathbf{v} \mp \mathrm{sign} \langle B_{r} \rangle \delta \mathbf{b}$ such that $\mathrm{sign} \langle B_{r} \rangle$ is the averaged polarity of the magnetic field. $<\cdot \cdot \cdot>$ represents a time average over a 60-minute non-overlapping time window which is based on the correlation times of Parker magnetic field fluctuations \citep{Parashar-2020, Chen-2020} and captures the full spectral features over the inertial range \citep{Ervin-2024SA}. 

% Further explanation on this can be found in \citet{Ervin-2024SA}.

\begin{equation}\label{eqn: sigmac}
    \sigma_C = \frac{<E^+> - <E^->}{<E^+> + <E^->}
\end{equation} 

We define slow wind to be {\alfic} (SASW) if every $|\sigma_C|$ value over a 30-minute time window is above 0.7, and considered classically slow (SSW) if it is below. This categorization method leads to three subcategories of wind based on the wind speed and {\sigmac} value. We identify streams of interest as time intervals lasting at least 30 minutes that fall into one of these categories for the entire stream. Due to the trajectory of Parker, the streams will traverse different longitudinal/spatial distances depending on the portion of the trajectory to which the stream corresponds. 

Along with the {\sigmac} calculation, we can also determine the residual energy (\sigmar), which describes the relative dominance between the energies associated with the velocity and magnetic field fluctuations, or can be thought of as the measure of alignment between the {\elas} variables. {\sigmar} $> 0$ ({\sigmar} $< 0$) is plasma dominated by energy in the fluctuating velocity (magnetic) field, and is calculated as follows:
% {\sigmar} is calculated via Equation~\ref{eqn: sigmar}  where 

\begin{equation}\label{eqn: sigmar}
    \sigma_R = \frac{2<\mathbf{z}^+ \cdot \mathbf{z}^->}{<E^+> + <E^->}
\end{equation} where $<\cdot \cdot \cdot>$ represents a time average over a 60-minute non-overlapping time window.

% For each type of wind we inspect data based on specific parameters, making sure we have at least a hundred consecutive data points before it can be labeled as a wind. However, instead of identifying each type of wind separately, as there are over 2 million data points total, we want to do it as efficiently as possible and perform such inspections in one go, resulting in around a 50\% performance increase versus identifying each wind type independently. The reason we can detect all three wind types simultaneously is because the designated parameters are mutually exclusive where each data point can belong only to a specific type of wind. As soon as at least a hundred consecutive data points are identified, we keep track of the starting index of that stream and keep adding points until there is a shift in wind (discontinuity in data), at which point, the identified indices are labeled with the appropriate wind type.

%%%%%%% -------------- Source Region Connection -------------- %%%%%%%
\subsection{Source Region Connection} \label{sec: methods-modeling} % [Tamar]
We use a combination of ballistic propagation \citep{Snyder-1966, Nolte-1973, Badman-2020, Macneil-2022, Koukras-2022} and potential field source surface (PFSS) modeling to connect the plasma observed at Parker to its estimated footpoint on the photosphere. The PFSS model is an efficient method to model the global coronal magnetic field out to an equipotential surface (source surface - {\Rss}) where the magnetic field is assumed to be purely radial. The model assumes a magnetostatic ($\nabla \times \mathbf{B} = 0$) corona, with a photospheric radial magnetic field observation as the inner boundary \citep{Schatten-1969, Altschuler-1969} to reconstruct the field out to the {\Rss}. We use the open source \texttt{sunkit-magex} {\footnote{https://github.com/sunpy/sunkit-magex}} package \citep{pfss} to reconstruct the coronal field out to a {\Rss} height of 2.5~{\Rsun} \citep{Hoeksema-1984} with ADAPT-GONG magnetograms as the inner boundary condition. We determine the best PFSS model for each encounter based on matching the observed neutral line (HCS) crossings at Parker with the modeled HCS from the PFSS instantiation \citep{Badman-2020, Ervin-2024CH}.

The PFSS model only reconstructs the magnetic field out to the source surface (2.5~{\Rsun}), and thus we need another method to connect the spacecraft trajectory which varies between 11~{\Rsun} and 40~{\Rsun} to the model outer boundary. Assuming a Parker spiral and varied wind speed (the in situ $v_R$ measurement), we propagate the spacecraft trajectory down to the {\Rss} height to connect our observations with the model. This allows us to connect the plasma observed at Parker with its estimated footpoint location at the photosphere and identify source regions based on comparison with EUV images and other parameters. 

In Appendix~\ref{appendix: validation} we discuss model validation and sources of error in both the ballistic propagation and PFSS model. We find that the error in the estimated footpoints due to ballistic propagation methods is velocity dependent (impacting the slower wind more) and is on the order of $1^{\circ}$. For the PFSS modeling results, we find that the source surface height has the largest impact on the resulting footpoints, however this can be optimized by comparing modeled polarity with magnetic field observations. The choice of input magnetogram leads to an error of $\sim 2^{\circ}$ when using magnetograms within a few days from the optimal input parameter. \citet{Neugebauer-2002} showed that expected uncertainties in PFSS source region footpoints are $\sim 10^{\circ}$ or less in longitude and while the error on the PFSS estimated footpoints is larger than for the ballistic propagation, the resulting source region is unaffected when using input parameters optimized to the observations leading to confidence in the results.

We use these modeling methods to closely examine {\saswnum} SASW streams and their sources in addition to the overall in situ characteristics of slow {\alfic} wind observed at Parker. We look at the source region and source characteristics of {\fswnum} FSW streams for comparison, as it has been well established that these streams emerge from CH regions and thus provide a good source of comparison for our modeled SASW streams \citep{McComas-1998, McComas-2008}.

% some definitions
In this paper, when we discuss coronal holes, we are referring to cool, low density regions of coronal plasma with rapidly expanding open field lines and low magnetic activity levels \citep{Cranmer-2009}. This manifests as relatively dark regions in EUV observations and low photospheric magnetic field values. When we refer to active regions (ARs), we are referencing regions of relatively high magnetic activity in the corona associated with strong photospheric magnetic fields in connection to bright loops in EUV \citep{Weber-2023}.

%%%%%%% ------------------------------------------------------------------------------------------------------------------------------ %%%%%%%
%%%%%%% -------------------------------------------------------- DISCUSSION -------------------------------------------------------- %%%%%%%
%%%%%%% ------------------------------------------------------------------------------------------------------------------------------ %%%%%%%
\section{Discussion} \label{sec: disc}

%%%%%%% -------------- In Situ Characteristics -------------- %%%%%%%
\subsection{In Situ Characteristics} \label{sec: disc-insitu} % [Kai]

\begin{figure} [ht] %[htb!]
\begin{center}
  \includegraphics[width=\columnwidth]{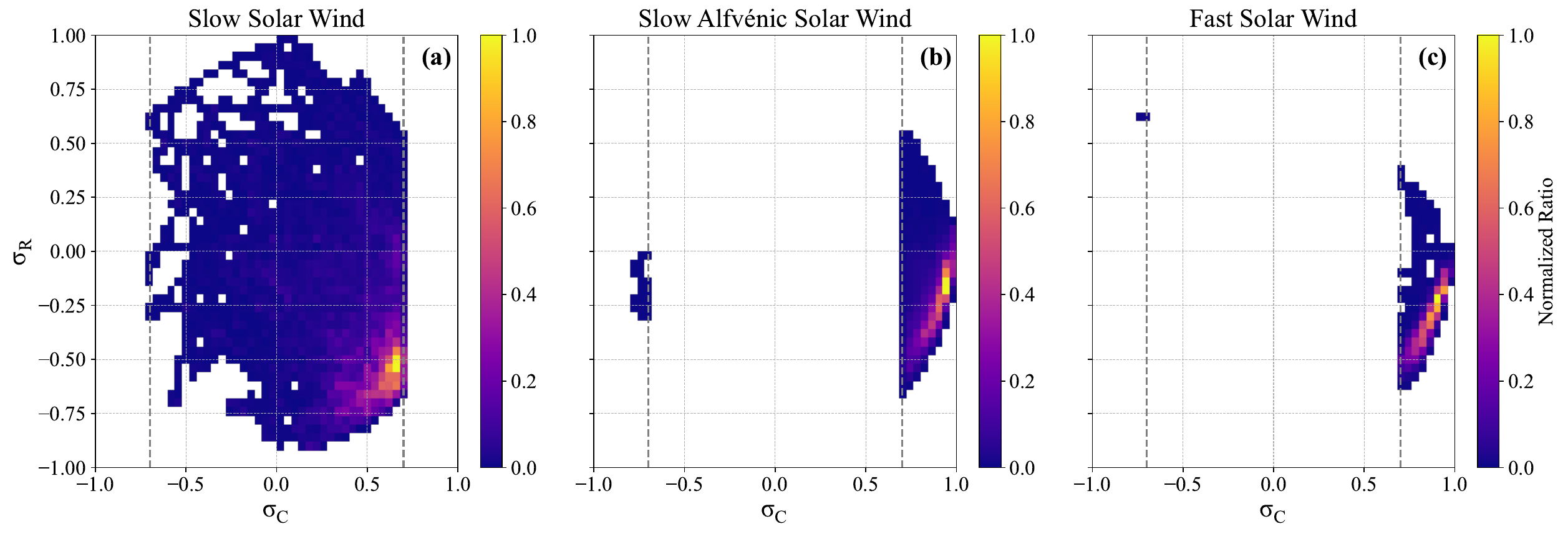}
  \caption{All usable E4 to E14 data (see Section ~\ref{sec: methods-identification}) is compiled into normalized histograms of cross helicity ({\sigmac}) against the residual energy ({\sigmar}) to showcase the identification method. The criteria for categorization of \lq{}high\rq{} {\alfty} and the radially dependent velocity thresholding method are described in Section~\ref{sec: methods-identification}. Panels (a), (b), and (c) plot the designated data for SSW, SASW, and FSW respectively.
  }
  \label{fig: Sigmac_Sigmar}
\end{center}
\end{figure}

% Residual energy vs cross helicity background
As discussed in Section~\ref{sec: intro}, the SASW shows high levels of {\alfty} yet slow wind speeds. Figure~\ref{fig: Sigmac_Sigmar} shows the relation of these cross helicity and residual energy (see Section~\ref{sec: methods-identification}) for different wind classifications.

% sigmac vs. sigmar
Due to the relationship between {\sigmac} and {\sigmar}, we expect $\sigma_C^2 + \sigma_R^2 \leq 1$ \citep{Bavassano-1998} meaning these measurements (as shown in Figure~\ref{fig: Sigmac_Sigmar}) are bounded by a circle. A comparison of the three panels, each for a different wind categorization, shows the stark differences in the turbulence parameters between the types of wind. From panel (a), it is clear that the SSW shows dramatically different properties than the SASW (panel (b)) and FSW (panel (c)). The SSW shows a large variance in {\sigmar} values, with the majority of the distribution at {\sigmar} $<$ 0, indicating the SSW is magnetically dominated and the {\sigmac} distribution is concentrated between 0.5 and 0.7. We expect the cross helicity to be more strongly concentrated at {\sigmac} $>$ 0, as the solar wind at these heliocentric distances is dominated by outward propagating modes.

The SASW and FSW distributions look quite similar: {\sigmar} $< 0$  and $|\sigma_C| \sim 1$, meaning that SASW and FSW are both magnetic energy dominant and the turbulence is strongly imbalanced: i.e. the energy fluxes in the inward and outward propagating wave modes ($E^{\pm}$) differ such that {\sigmac} $\neq 0$ \citep{Sioulas-2024}. The turbulence and energy properties of the SASW are far more similar to the FSW than the SSW, potentially indicating a similar source region and/or acceleration mechanism as the FSW \citep{DAmicis-2015, DAmicis-2018, Stansby-2020alf, Ervin-2024CH}. 

% The impact of imbalance on higher-order turbulence statistics is an open area of investigation. Recent work by \citet{Shi-2023, Sioulas-2024, Sioulas-2024SDDA} and others have looked at the scaling of alignment angles, spectral indices, and the impact on the principles of \lq{}critical balance\rq{} \citep{Goldreich-1995, Goldreich-1997} and \lq{}scale-dependent dynamic alignment\rq{} \citep{Boldyrev-2006, Chandran-2015, Mallet-2017} based on the level of imbalance.

% We can see that regardless of the wind speed, this subcategory of SASW has similar properties of FSW and looks much different than the SSW with low {\sigmac} and {\sigmar} values.

% is expected that  cross helicity ({\sigmac}) is plotted against the residual energy ({\sigmar}), a circular shape is expected and is seen in Figure~\ref{fig: Sigmac_Sigmar}, which is a validation of the cross helicity calculation we used to determine {\alfty} of the plasma. is much different for which energy is dominant in the plasma compared to the other plots. 

%%% FIGURE --- Sigmac vs Sigmar

% Helium Abundance Background
% Now that we have a better understanding of the relationship between {\alf} wave propagation and magnetic field fluctuations, we can now study the composition of the solar wind which is useful in providing insight in identifying their source regions.
Classifying the solar wind into these categories allows us to study the sources and compositional metrics (alpha-to-proton abundance) of these wind types. In this study that focuses on solar minimum and the rising phase, high {\Ahe} will likely be associated with open field line or CH-like structures where fast speed wind emanates from, however this is not always the case (see Section~\ref{sec: intro}). 

% The alpha particle (Helium) abundance is the ratio of alpha particle and proton densities ($\mathrm{A_{He} = N_\alpha / N_p}$). Elemental composition is often used as a tracer of source region due to the conservation of plasma composition throughout the solar wind's radial evolution. Therefore, this can provide a helpful metric to compare between solar wind types in association with modeled source region mapping results. Generally, alpha particle abundances are shown to be solar cycle dependent in slow solar wind, ranging between \Ahe$\sim1-4\%$, while abundances become less and less solar cycle dependent with increasing solar wind speed, remaining consistently around \Ahe$\sim4\%$ \citep{Kasper-2007}. In particular, He abundances exhibit the largest difference during periods around solar minimum/rising phase, where faster speed wind is observed to contain \Ahe$\sim4\%$ while the slowest speed observed near the Earth contains 1\%, making solar wind types the most distinguishable during this phase of the solar cycle. In general, alpha abundances are depleted compared with photospheric values, however, exhibit changes across solar wind values and thus can be used for analysis. Therefore, in this study that focuses on solar minimum and the rising phase, high {\Ahe} will likely be associated with open field line or CH-like structures where fast speed wind emanates from, however this is not always the case (see Section~\ref{sec: intro}). 

\begin{figure} [ht] %[htb!]
\begin{center}
  \includegraphics[width=\columnwidth]{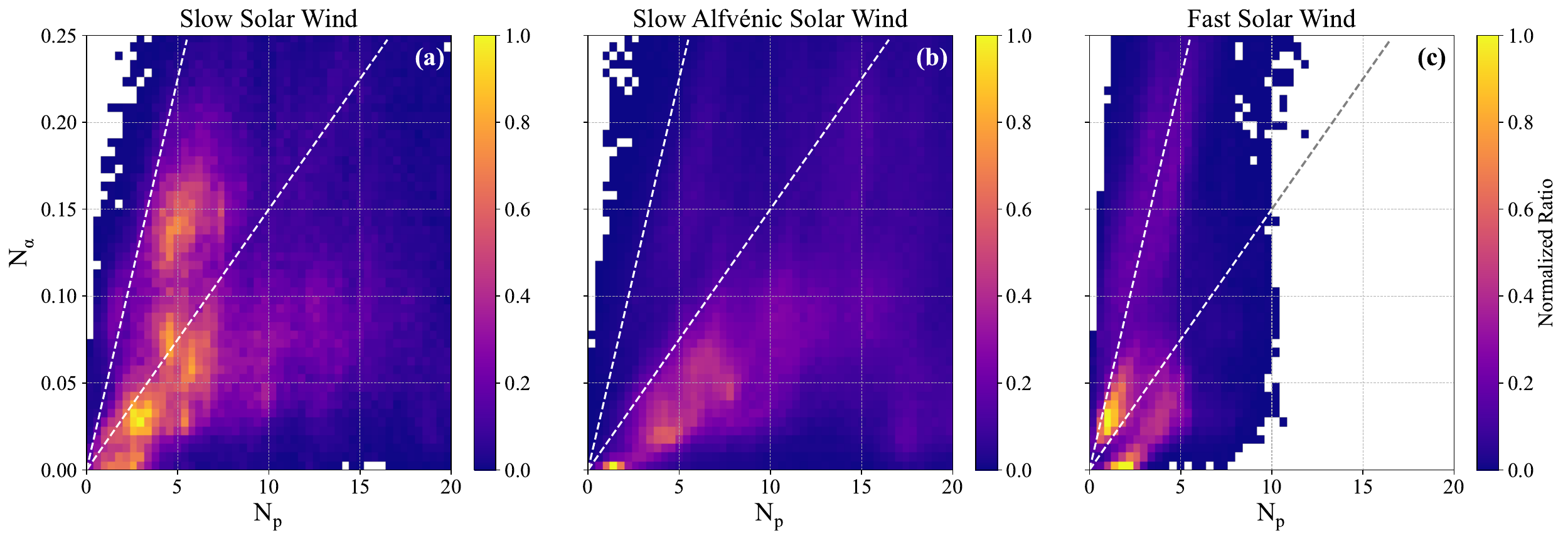}
  \caption{All usable E4 to E14 data (see Section ~\ref{sec: methods-identification}) is compiled into these normalized histograms of alpha particle density ($N_\alpha$) against proton density ($N_p$), each scaled by $R^2$. Particle measurements from the SPAN-I instrument aboard Parker/SWEAP show the alpha-to-proton abundance ratios. The dashed lines indicate the typical \lq{}high helium abundance\rq{} threshold of 0.045 and the \lq{}low abundance\rq{} boundary of 0.015 \citep{Kasper-2007, Kasper-2012}. Panels (a), (b), and (c) plot the designated data for SSW, SASW, and FSW respectively.
  }
  \label{fig: helium_abundance}
\end{center}
\end{figure}

% helium abundance overview
Figure~\ref{fig: helium_abundance} shows the proton and alpha particle composition for each wind type, where panel (a) looks at the SSW, panel (b) the SASW, and panel (c) the FSW. We include dashed lines at {\Ahe} = 0.015 and 0.045 to indicate the \lq{}low\rq{} and \lq{}high\rq{} abundance thresholds \citep{Kasper-2007, Kasper-2012} and to guide the eye.

We see that the {\Ahe} for the SSW and SASW rarely cross the typical \lq{}high abundance\rq{} threshold of 0.045 and that the SASW distribution is largely concentrated below the \lq{}low abundance\rq{} boundary of 0.015. The SSW shows both intermediate and low alpha particle abundance. The FSW distribution clusters around the \lq{}high abundance\rq{} threshold of 0.045 (as to be expected from CH wind), with some anomalous fast wind showing lower alpha densities than expected.

% The SASW appears to show two populations of low {\Ahe} plasma, one with higher proton densities than the other, potentially indicating that these populations originate from different source regions: the high {\Ahe} population from CH-like structures, and the low abundance population from closed loop or streamer structures.

%%% FIGURE --- helium abundance

While the categorization of the slow wind by cross helicity is useful to separate populations from potentially different sources, including compositional characteristics in categorization schemes would be a useful method to better categorize the wind by its source region. Additional observations of heavy ions and molecular composition would aid in this process allowing for better connection between in situ observations and sources, as our understanding of alpha particle abundance and its relation to source region is still convoluted and dependent on phase in the solar cycle.

% These results explain why it is necessary to have another categorization of solar wind, SASW, due to its unique properties in helium abundance and specifically in its curiously large density of protons.

% Conclusion
The data highlighted above provides a strong rationale to study this additional categorization of the SASW due to its unique properties in energy and elemental composition in comparison to more classical fast and slow wind. The SASW is seen to have similar properties to the FSW in terms of {\sigmac} and {\sigmar}, supporting the evidence that at least some of SASW emerges from similar structures to the FSW. However, it is noteworthy that the observed alpha abundance for the SASW is much lower than that of the FSW and the observed proton abundance is larger than the FSW (indicating additional mass flux), implying that some of this plasma likely has a unique source region or acceleration mechanisms, warranting additional study. 

%%%%%%% -------------- Source Region Characteristics -------------- %%%%%%%
\subsection{Source Region Characteristics} \label{sec: disc-modeling} % [Tamar]
We use the PFSS and ballistic propagation methods to compare results produced from the \lq{}estimated\rq{} footpoints for the SASW to the types of wind. We can calculate the expansion factor, photospheric magnetic field value at the footpoint, and brightness in EUV. Due to the inherent uncertainties in solar wind source modeling, doing this for one stream can be inconclusive and therefore we look at a variety of streams from E4 to E14 to have a more representative look at the potential sources of the SASW. In Tables~\ref{tab:sasw-overview} and~\ref{tab:fsw-overview}, we show an overview of the SASW and FSW streams we model and some of their associated characteristics. 

\begin{table}[]
    \centering
    \begin{tabular}{|c|c|c|c|c|c|c|c|c|}
    \hline
    \textbf{Time Period} & \multirow{1}{*}{\textbf{$\mathrm{\langle v_r \rangle}$}} & \multirow{1}{*}{\textbf{$\mathrm{\langle |\sigma_C| \rangle}$}} & \multirow{1}{*}{\textbf{$\mathrm{\langle M_A \rangle}$}} & \multirow{1}{*}{\textbf{$\mathrm{\langle B_r R^2 \rangle}$}} & \multirow{1}{*}{\textbf{$\mathrm{\langle N_p R^2 \rangle}$}} & \multirow{1}{*}{$\mathrm{\langle B_0 \rangle}$} & \multirow{1}{*}{$\mathrm{\langle f_{ss} \rangle}$} & \multirow{1}{*}{$\mathrm{\langle I \rangle}$} \\
    & [{\kms}] & &  & [nT AU$^2$] & [cm$^{-3}$] & [G] & & \\ [3pt]
    \hline 
    2020-01-28 18:00:00 to 2020-01-29 01:00:00 (E4) & 284 & 0.86 & 2.06 & -1.95 & 7.47 & -4.27 & 236 & 30.9 \\
    \hline 
    2020-06-06 06:00:00 to 2020-06-06 09:00:00 (E5) & 306 & 0.87 & 2.19 & -2.03 & 6.77 & -1.55 & 31.7 & 7.68 \\
    \hline 
    2020-06-09 03:00:00 to 2020-06-09 22:00:00 (E5) & 217 & -0.92 & 2.45 & 2.01 & 12.7 & 2.28 & 27.5 & 11.4 \\
    \hline
    2020-09-27 12:00:00 to 2020-09-27 16:00:00 (E6) & 306 & -0.92 & 1.51 & -2.12 & 7.54 & -3.13 & 48.2 & 10.3 \\
    % \hline
    % 2020-09-29 02:00:00 to 2020-09-29 05:00:00 (E6) & 369 & 0.86 & 1.60 & -2.58 & 4.20 & -2.98 & 48.2 & 85.1 \\
    \hline
    2021-01-18 18:00:00 to 2021-01-19 06:00:00 (E7) & 231 & 0.86 & 2.29 & -2.58 & 4.20 & -5.86 & 27.7 & 23.6 \\
    \hline
    2021-04-30 00:00:00 to 2021-04-30 05:00:00 (E8) & 155 & 0.86 & 1.23 & -1.99 & 20.6 & 37.8 & 105.7 & 395 \\
    \hline
    2021-04-30 10:00:00 to 2021-05-01 00:00:00 (E8) & 192 & -0.95 & 1.70 & -1.99 & 18.7 & 3.43 & 11.4 & 96.5 \\
    \hline
    2022-12-10 18:00:00 to 2022-12-10 19:00:00 (E14) & 187 & -0.92 & 0.94 & -2.65 & 17.9 & -118 & 354 & 64.9 \\
    \hline
    2022-12-10 20:30:00 to 2022-12-10 22:00:00 (E14) & 178 & -0.92 & 0.79 & -2.81 & 16.7 & -146 & 4798 & 596 \\
    \hline
    2022-12-12 12:00:00 to 2022-12-12 16:00:00 (E14) & 240 & -0.90 & 1.36 & 2.56 & 16.6 & -16.8 & 39 & 316 \\
    \hline
    \end{tabular}
    \caption{Overview of the {\saswnum} SASW streams identified and modeled in this study. The $\langle \cdot \cdot \cdot \rangle$ refers to an average over these parameters for the entire stream. $\mathrm{v_r}$, $\mathrm{\sigma_C}$, {\MA}, $\mathrm{B_r R^2}$ and $\mathrm{N_p R^2}$ are parameters derived from in situ Parker measurements, while $\mathrm{B_0}$, $\mathrm{f_{ss}}$, and $\mathrm{I}$ are determined from the model estimated footpoints of the streams. $\mathrm{B_0}$ is the photospheric footpoint field strength, $\mathrm{f_{ss}}$ is the magnetic expansion factor as defined by Equation~\ref{eqn: fss}, and $\mathrm{I}$ is the intensity of the footpoint from the corresponding AIA 193~{\AA} image.}
    \label{tab:sasw-overview}
\end{table}

\begin{table}[]
    \centering
    \begin{tabular}{|c|c|c|c|c|c|c|c|c|}
    \hline
    \textbf{Time Period} & \multirow{1}{*}{\textbf{$\mathrm{\langle v_r \rangle}$}} & \multirow{1}{*}{\textbf{$\mathrm{\langle |\sigma_C| \rangle}$}} & \multirow{1}{*}{\textbf{$\mathrm{\langle M_A \rangle}$}} & \multirow{1}{*}{\textbf{$\mathrm{\langle B_r R^2 \rangle}$}} & \multirow{1}{*}{\textbf{$\mathrm{\langle N_p R^2 \rangle}$}} & \multirow{1}{*}{$\mathrm{\langle B_0\rangle}$} & \multirow{1}{*}{$\mathrm{\langle f_{ss} \rangle}$} & \multirow{1}{*}{$\mathrm{\langle I \rangle}$} \\
    & [~{\kms}] & &  & [nT AU$^2$] & [cm$^{-3}$] & [G] & & \\ [3pt]
    \hline 
    2020-01-27 04:55:00 to 2020-01-27 05:15:00 (E4) & 412 & 0.86 & 2.30 & -2.18 & 3.78 & -5.43 & 34.6 & 14.5 \\
    \hline 
    2021-04-27 01:00:00 to 2021-04-27 04:00:00 (E8) & 436 & 0.86 & 1.54 & -2.51 & 2.70 & -15.1 & 75.5 & 81.7 \\
    \hline 
    2021-04-27 06:00:00 to 2021-04-27 08:30:00 (E8) & 443 & 0.85 & 1.67 & -2.45 & 3.25 & -15.0 & 75.8 & 88.6 \\
    \hline
    2021-11-20 05:30:00 to 2021-11-20 10:30:00 (E10) & 563 & 0.88 & 1.48 & -2.10 & 2.91 & -12.9 & 34.4 & 34.8 \\
    \hline
    2022-12-15 21:15:00 to 2022-12-15 22:15:00 (E14) & 473 & -0.80 & 2.21 & 3.47 & 3.36 & 17.0 & 15.9 & 126.7 \\
    \hline
    \end{tabular}
    \caption{Overview of the {\fswnum} FSW streams identified and modeled in this study. The $\langle \cdot \cdot \cdot \rangle$ refers to an average over these parameters for the entire stream. $\mathrm{v_r}$, $\mathrm{\sigma_C}$, {\MA}, $\mathrm{B_r R^2}$ and $\mathrm{N_p R^2}$ are parameters derived from in situ Parker measurements, while $\mathrm{B_0}$, $\mathrm{f_{ss}}$, and $\mathrm{I}$ are determined from the model estimated footpoints of the streams. $\mathrm{B_0}$ is the photospheric footpoint field strength, $\mathrm{f_{ss}}$ is the magnetic expansion factor as defined by Equation~\ref{eqn: fss}, and $\mathrm{I}$ is the intensity of the footpoint from the corresponding AIA 193~{\AA} image.}
    \label{tab:fsw-overview}
\end{table}

%%%% --- footpoint field strength

\begin{figure} [ht] %[htb!]
\begin{center}
  \includegraphics[width=\columnwidth]{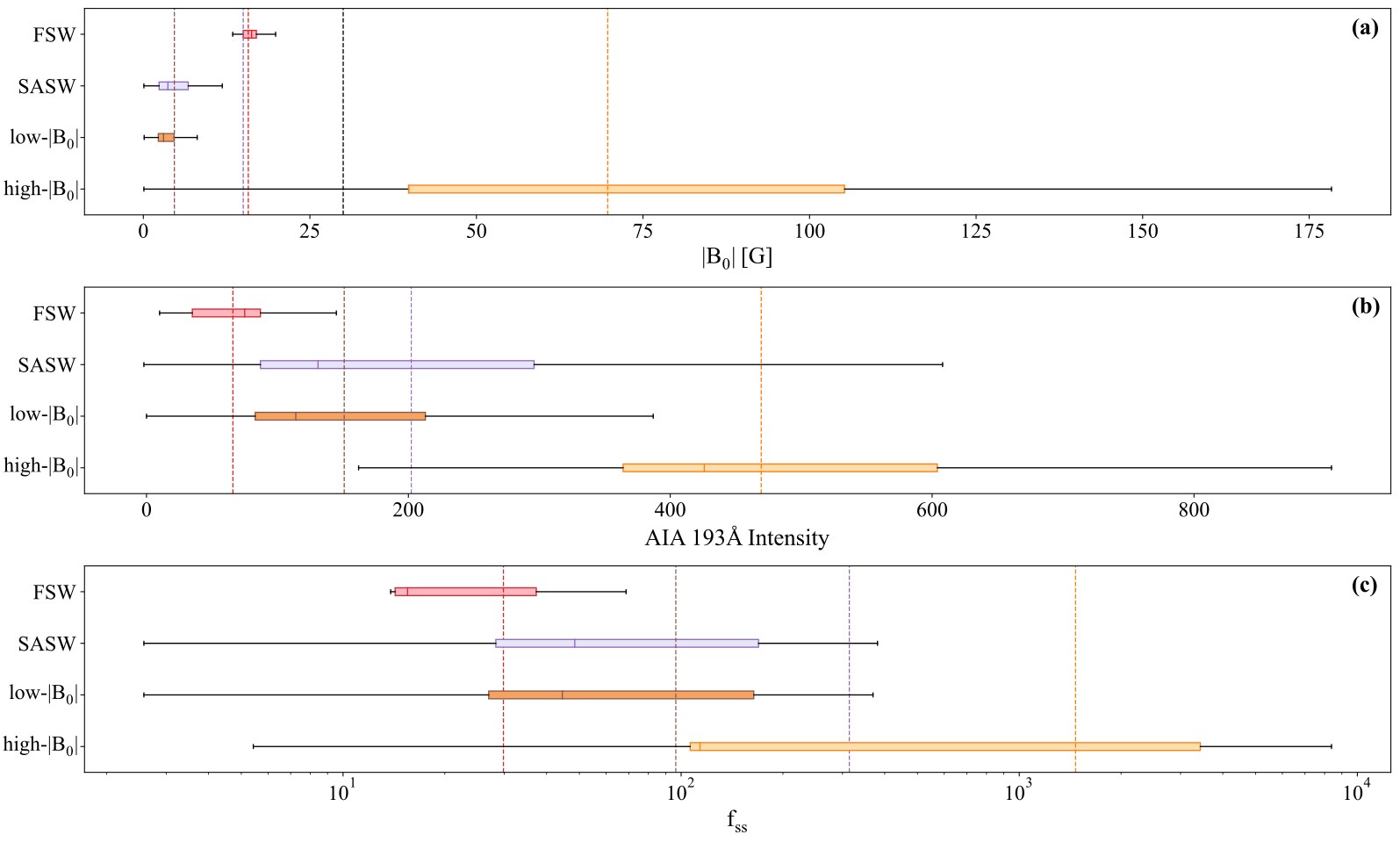}
  \caption{Overview of parameters determined from PFSS modeling and the associated estimated footpoints for the FSW. SASW, and the CH-like and non-CH SASW sub-categories. The boxplot shows the inner quartile range (25 - 75\%) percentile and the solid line shows the median value. The dashed vertical lines show the mean of each parameter for the type of wind stream. Panels (a), (b), and (c) show the photospheric footpoint field strength ({\Bo}), intensity at the footpoint from AIA 193 ~{\AA} images, and expansion factor based on Equation~\ref{eqn: fss} respectively.}
  \label{fig: footpoint_expansion}
\end{center}
\end{figure}

Based on the footpoint estimations, we can determine $B_0$ or the photospheric magnetic field value at the estimated point of plasma emergence by sampling GONG magnetograms at the footpoint locations. \citet{Stansby-2021aa} studied the contribution of active regions to the solar wind over four solar cycles using GONG magnetograms (the same magnetograms used in this study), setting a lower limit on the field strength of ARs by looking at the maximum magnetic field strength outside of AR latitudes ($\pm$30{\degree} in the corona and $\pm$60{\degree} in the solar wind), finding it to be 30 G for GONG, well above the established noise level of 3G \citep{Clark-2003}. In Panel (a) of Figure~\ref{fig: footpoint_expansion}, we look at the photospheric magnetic field strength at the footpoints for wind identified as fast and slow {\alfic} as described in Section~\ref{sec: methods-identification}.

We find that the FSW streams show footpoint field strengths well below the 30 G threshold, as expected from wind emerging from CH-like structures. The overall distribution of modeled SASW streams also show footpoint field strengths below 30 G, however when we look at the average values across individual streams (Table~\ref{tab:sasw-overview}), we find that there are some streams where the average value exceeds (sometimes even dramatically) this threshold. We then further sub-divide our SASW streams into low-{\Bo} streams ($\langle \mathrm{B_0} \rangle <$ 30 G) and high-{\Bo} streams ($\langle \mathrm{B_0} \rangle >$ 30 G). With this reclassification, we now see a dramatic difference in the overall spread in the photospheric field strength, specifically the three \lq{}high-{\Bo}\rq{} streams show {\Bo} values significantly higher than the \citet{Stansby-2021aa} threshold.

%%% FIGURE --- footpoint field strength & expansion factor

\begin{figure} [ht] %[htb!]
\begin{center}
  \includegraphics[width=\columnwidth]{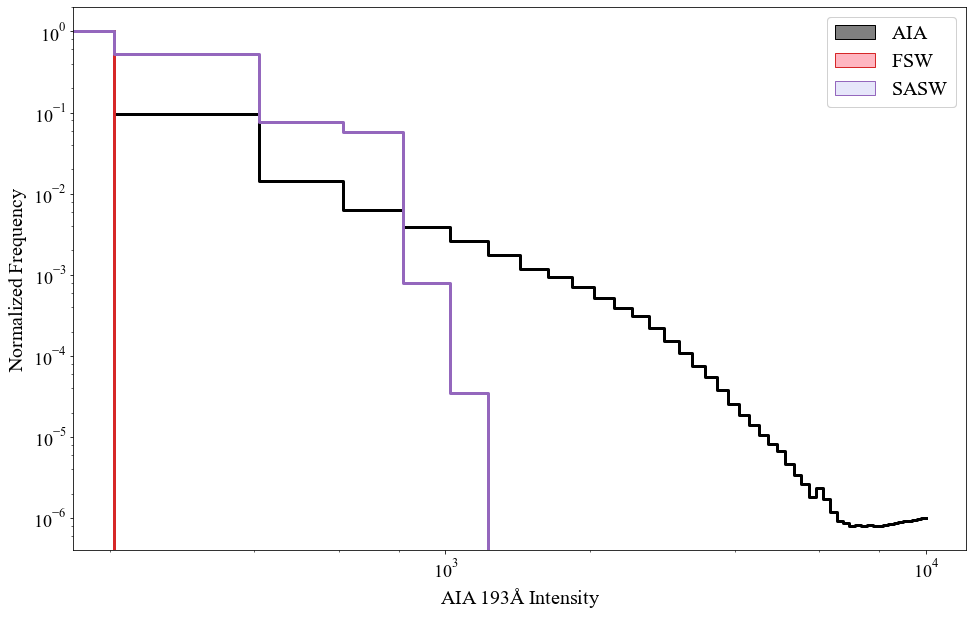}
  \caption{Normalized histograms comparing intensity values from many AIA images with intensity values sampled at the footpoints of the {\saswnum} identified and modeled SASW streams. The AIA images chosen are images at a 24-hour cadence for two weeks around each observation.}
  \label{fig: intensity_hist}
\end{center}
\end{figure}

%%%% --- intensity
Identical to our methods of sampling the GONG magnetograms at the estimated footpoint to determine {\Bo}, we can sample EUV observations (SDO/AIA 193~{\AA} images) at these same footpoints to extract \lq{}intensity\rq{} values. Brightness in EUV channels is dependent on the electron temperature and density, therefore hotter/denser ARs are brighter and cooler/tenuous CH structures are darker. We choose EUV images for comparison based on the estimated time the plasma left the photosphere which is determined based on the speed of the stream of interest and position of Parker. In Figure~\ref{fig: intensity_hist}, we compare EUV intensity at the estimated footpoints of the {\saswnum} SASW streams we studied in depth, with the distribution of intensity values from a variety of AIA images corresponding to the two weeks around each of the studied SASW streams to give an overview of \lq{}typical\rq{} distributions of intensity values. We find that the distribution of intensity values typically associated with SASW streams falls on the \lq{}darker\rq{} (cooler) end of the distribution when compared with intensity values across many AIA images. Additionally, from Panel (b) of Figure~\ref{fig: footpoint_expansion}, we see that the high-{\Bo} SASW shows much higher intensity values in comparison with the low-{\Bo} and FSW, likely indicating a non-CH source.

%%%% FIGURE --- intensity

%%%% --- expansion factor
We calculate the magnetic expansion factor ($f_{ss}$) between the photosphere ($R_0 =$ 1.0~{\Rsun}) and another radial distance ($R_1$) based on \citet{Wang-1990, Wang-1997}:
\begin{equation}\label{eqn: fss}
    f_{s} = \left( \frac{R_0}{R_1}\right)^2 \frac{B_R(R_0, \lambda_0, \phi_0)}{B_R(R_1, \lambda_1, \phi_1)}
\end{equation}
such that expansion factor describes the divergence of the flux tube from spherical expansion ($f_{s} = 1$) meaning values of $f_{s}$ greater than 1 indicate a rapidly diverging flux tube \citep{Wang-1997}. In this case, $R_1$ = {\Rss}, meaning our expansion factor is a measure of the expansion to the source surface: $f_{ss}$, which is the quantity previously correlation with heliospheric wind speed \citep{Wang-1990, Riley-2012, Riley-2015}. 

% From the PFSS method, we can also calculate the magnetic expansion factor, where a larger expansion factor implies more rapid divergence from spherical expansion ($f_{ss} = 1$). 
% As discussed in the introduction, we would expect higher magnetic expansion factors from slow wind emerging from the same regions as fast wind, as rapid divergence of the flux tube leads to a higher critical point and thus the majority of the heating occurs below the transition to supersonic wind speeds \citep{Leer-1980}. 
In Panel (c) of Figure~\ref{fig: footpoint_expansion}, we find that in general the SASW has a higher expansion factor than the FSW. With our additional classification into high-{\Bo} and low-{\Bo} SASW, we find that the high-{\Bo} SASW has expansion factors an order of magnitude larger than the low-{\Bo} SASW and FSW, indicating a rapidly diverging field with a higher critical point, meaning more heating below the critical point leading to slower wind speeds. Compared with the FSW, the low-{\Bo} SASW has larger expansion, indicating that this is likely wind emerging from an overexpanded open field line in a CH-like region with additional heating below the critical point leading to slower wind speeds, but from a similar structure as the FSW.

While anti-correlation between expansion factor and wind speed is expected \citep{Wang-1990} and informs many current solar wind models such as the Wang-Sheeley-Arge (WSA) Model \citep{Arge-2000, Arge-2003, Arge-2004} there is not a one-to-one correlation between speed and {\fss}. Other models, such as the distance from the coronal hole boundary (DCHB) model \citep{Riley-2015} depend on the angular distance from a CH boundary to determine the solar wind speed, rather than the expansion factor \citep{Riley-2001}. This indicates that our understanding of the expansion and acceleration processes in the solar wind are incomplete, and relying solely on the expansion factor to infer a final wind speed is not a fully accurate description of the system.

% \citep{Dakeyo-2024Expansion}

Our observations of high expansion factors associated with high EUV intensity and strong footpoint magnetic field strength of these \lq{}high-{\Bo}\rq{} streams alongside the long lived fan structures seen in EUV observations (see Appendix~\ref{appendix: streams}) the idea of \citet{Yardley-2024}. This would allow for plasma with non-CH composition to be released from coronal loop like structures after reconnection with open field lines leading to in situ composition observations different from the FSW \citep{Fisk-2003}. If we had measurements of heavy ion ratios and FIP bias from Parker, we would expect to see enhanced heavy ion ratios ({\oxy}, {\car}) and {\feo} ratios above typical photospheric values.

%%%% FIGURE --- expansion factor 
% \begin{figure} [ht] %[htb!]
% \begin{center}
%   \includegraphics[width=\columnwidth]{expansion_factor.png}
%   \caption{A comparison of the source region characteristics as derived from the PFSS modeling results for the different wind streams as described in Section~\ref{sec: methods-modeling}. Panels (a), (b), and (c) show results for the FSW, SASW, and SSW streams identified respectively.}
%   \label{fig: expansion_factor}
% \end{center}
% \end{figure}

\subsection{In Situ Characteristics Associated with Modeled Streams} \label{sec: disc-comp-model}
\begin{figure} [ht] %[htb!]
\begin{center}
  \includegraphics[width=\columnwidth]{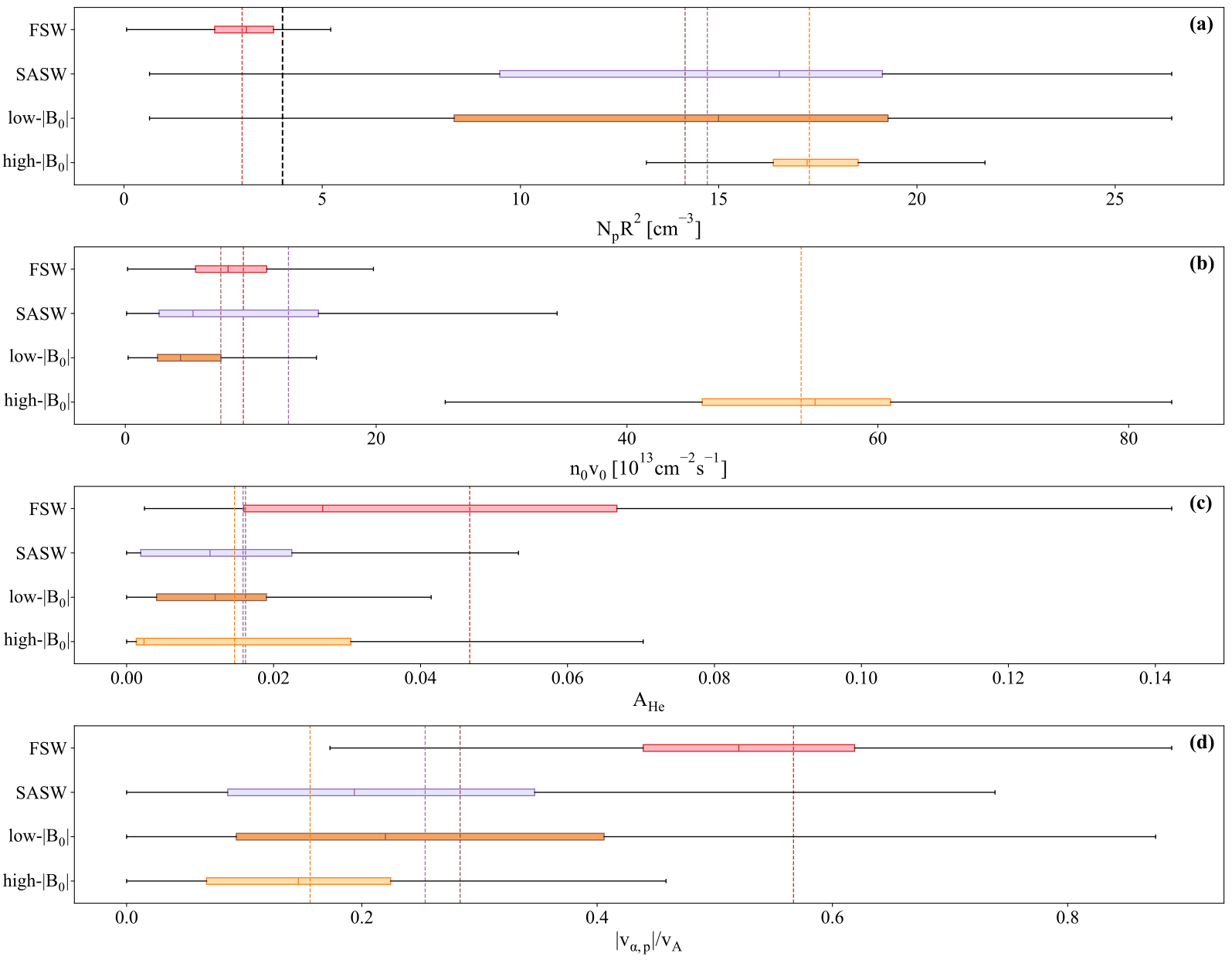}
  \caption{Box plot overviews of compositional parameters for the {\saswnum} SASW and {\fswnum} FSW streams we modeled. The dashed vertical lines indicate the mean for the distribution with the corresponding color and the solid line shows the median. 
  \textit{Panel (a):} Scaled proton density from SWEAP/SPAN-I. The dashed black line indicates typical 1 AU solar wind densities.
  \textit{Panel (b):} Proton mass flux ($n_0 v_0 = (B_0 / B_r) N_P v_P$) calculated from in situ observations and the photospheric footpoint field strength from modeling results.
  \textit{Panel (c):} Alpha abundance ratio ({\Ahe}) using alpha and proton measurements from SWEAP/SPAN-I. 
  \textit{Panel (d):} Normalized differential streaming speed ({\vap} / $\mathrm{v_A}$). 
  }
  \label{fig: composition}
\end{center}
\end{figure}

%%%% --- composition
In Section~\ref{sec: disc-insitu}, we looked at the overall distribution of proton and alpha particles from Parker observations. Now we will look at the composition of the modeled FSW and SASW streams in more detail to compare in situ metrics with potential sources. In Figure~\ref{fig: composition}, we show an overview of the scaled proton density ($\mathrm{N_p R^2}$), alpha particle abundance ({\Ahe} = $N_p / N_{\alpha}$), mass flux density ($n_0 v_0$), and normalized differential streaming speed ({\vap} / $\mathrm{v_A}$) where {\vap}$\mathrm{= (v_{\alpha} - v_p)/ \cos{\theta}}$. $\cos{\theta} = |B_R/B|$ to account for magnetic polarity, and $\mathrm{v_{\alpha}}$, $\mathrm{v_p}$ are the alpha particle and proton radial velocities. We find that the FSW streams have significantly lower proton densities and relatively higher {\Ahe} than the SASW. The low-{\Bo} SASW shows a large range in proton density and lower alpha particle abundance than the high-{\Bo} streams, however both have mean values that fall within the \lq{}intermediate\rq{} abundance range (0.015 to 0.045 based on \citet{Kasper-2007, Kasper-2012}). This is similar to results for SASW previously observed at Parker by \citet{Huang-2020, Ervin-2024CH, Rivera-2024} and others, contributing to the confusion over source region of the SASW as there is not one abundance level that corresponds to this wind type. The statistical separation of the in situ data (which is independent from our modeling results) bolsters confidence that the source mapping provides meaningful information, and indicates that there might be multiple source regions for the SASW, some similar to FSW sources (CHs and their boundaries) and some different. 

If we assume a frozen-in magnetic field, mass will be conserved along a flux tube allowing for calculation of the proton mass density and energy flux at the coronal base \citep{Wang-2003}. The proton mass flux $n_0 v_0$ is calculated from the photospheric footpoint field strength determined from the modeling alongside in situ parameters: $n_0 v_0 = (B_0 / B_r) N_P  v_P$. In Figure~\ref{fig: composition}, we show the calculated proton mass flux, and find that the high-{\Bo} wind shows dramatically higher mass flux than the low-{\Bo} SASW and FSW (similar to the results from \citet{Wang-2003}). Higher mass flux is indicative of more heating below the critical point (near the coronal base), which when coupled with the high expansion that increases the height of the critical point, is expected to produce wind with slower speeds \citep{Wang-2012}. The low-{\Bo} SASW shows lower mass flux, similar to the FSW. This is thought to be regulated by the footpoint field strength \citep{Wang-2009}, and associated with temperature of the coronal source \citep{Stansby-2021massflux}. Therefore, low mass flux SASW should be associated with low footpoint field strength and cooler coronal regions, or CH-like structures. The higher expansion factor of the low-{\Bo} SASW would allow for slower wind to emerge.

% As shown in Figure~\ref{fig: composition}, the high-{\Bo} wind has significantly larger mass flux which when coupled with the high expansion indicating a rapidly diverging field with a higher critical point, means more heating below the critical point leading to slower wind speeds. The low-{\Bo} SASW has similar mass flux to the FSW, but larger expansion, indicating that this is likely wind emerging from an overexpanded open field line in a CH-like region with additional heating below the critical point leading to slower wind speeds, but from a similar structure as the FSW.

% The CH-like SASW shows slightly higher mass fluxes than the FSW, indicating that there is likely heating below the critical point leading to slower wind speeds.

%%%% FIGURE --- composition 

In Figure~\ref{fig: composition} Panel (d), we show the normalized differential streaming speed ({\vap} / $\mathrm{v_A}$) and see that the FSW streams show much higher differential speeds. \citet{Mostafavi-2022} showed correlation between {\vap} and {\vsw} and we note that the low-{\Bo} SASW has higher differential speeds than the high-{\Bo} SASW. This could point to a similar mechanism of preferential acceleration of alpha particles between the FSW and low-{\Bo} SASW which would follow from these streams having similar sources.

% Understanding this preferential acceleration of alphas remains an open question, and more work to address this 
%%%%%%% ------------------------------------------------------------------------------------------------------------------------------ %%%%%%%
%%%%%%% -------------------------------------------------------- RESULTS -------------------------------------------------------- %%%%%%%
%%%%%%% ------------------------------------------------------------------------------------------------------------------------------ %%%%%%%
% \section{Results} \label{sec: results}
% \begin{enumerate}
%     \item We find that the slow and slow {\alfic} solar wind typically show lower helium abundance in comparison with the fast wind.
% \end{enumerate}

%%%%%%% ------------------------------------------------------------------------------------------------------------------------------ %%%%%%%
%%%%%%% -------------------------------------------------------- CONCLUSION -------------------------------------------------------- %%%%%%%
%%%%%%% ------------------------------------------------------------------------------------------------------------------------------ %%%%%%%
\section{Results and Conclusions} \label{sec: conclusion}

%%%%%%% -------------- Results -------------- %%%%%%%
Using a wind speed classification model based on distance from the Sun and absolute cross helicity of the stream, we study the characteristics and parameters derived from in situ observations and modeling the footpoints of slow {\alfic} wind streams. We find that some (if not many) of these streams originate from coronal holes and their boundary regions.

\begin{enumerate}

    \item Using a heliocentric distance based classification scheme, we find distinct in-situ signatures of fast, slow, and slow {\alfic} solar wind that allow for categorization of wind streams as one of these three types. The SASW shows similar turbulence properties as the FSW and a large variance in alpha particle abundances which may point to a combination of sources.

    \item Through PFSS modeling, we find a population of low-{\Bo} slow {\alfic} wind where the footpoint field strength is below the typical threshold for active regions (typically below 10 G). These low-{\Bo} streams are characterized by high levels of {\alfty}, slow wind speeds, and intermediate alpha particle abundance in situ. From modeling we find they show a high expansion factor, low mass flux, and low intensity in EUV (associated with cold temperatures) indicating large heating below the critical point of a cooler structure. These streams show higher differential speeds indicating preferential acceleration of alpha particles. This is indicative of slow wind emerging from (overexpanded) CH-like or CH boundary structures.
    
    \item A second population of SASW shows high-{\Bo} (up to 100s G) alongside a even higher expansion factor, high mass flux, and high footpoint intensity indicating large heating near the coronal base in a hotter structure. Their alpha particle abundance is relatively low and they are associated with incredibly low differential speeds. When compared with EUV observations, these wind streams look to be perhaps emerging from the periphery of active regions or other closed loop structures. These high-{\Bo} streams likely emerge from non-CH regions, perhaps from closed loop regions through interchange reconnection, however additional work using heavy ion and elemental composition metrics and other observations is required for conclusive results.
    
\end{enumerate}

While this work provides an overview of some slow {\alfic} solar wind streams observed by Parker, their modeled sources and associated in situ parameters, there are many caveats that must be considered when coming to conclusions about the true source of the SASW due to the complexity associated with modeling its origins. In this work, we use the potential field source surface (PFSS) combined with ballistic propagation to estimate the footpoints associated with wind observed in situ by Parker. This modeling method makes huge (potentially unrealistic) assumptions about the magnetic topology of the corona. Additionally, these modeling methods can only trace to sources connected to field lines that are topologically \lq{}open\rq{} at the source surface, meaning that connecting to wind from the periphery of closed loops, or streamer blobs is difficult with both aspects likely requiring time dependent modeling. While PFSS can be validated, it is important to remember that these are all \lq{}estimated\rq{} footpoints and additional work such as looking at elemental and charge state composition, and comparison with other modeling methods such as magnetohydrodynamic (MHD) models to better capture the more complicated physics in the corona would be useful to have additional confidence in tracing of sources of the slow wind. However, we note that the statistical approach we take here reveals robust trends in source region properties. In particular, it is non-trivial that the properties inferred from modeling and sampling remote observations are systematically distinct when filtering wind of the same speed by cross helicity, an in situ property. 

%%%%%%% -------------- Future Work -------------- %%%%%%%
% To fully understand and resolve the likely multitude of source regions associated with the various types of slow solar wind, new data sets, ideally combining in situ and remote observations will need to be produced. 
Solar Orbiter, through the Heavy Ion Sensor (HIS; \citet{Livi-2023}) will allow us to study the compositional characteristics of the wind in situ and compare with its source region using Spectral Imaging of the Coronal Environment (SPICE; \citet{Anderson-2020}) instrument. Studies using the combination of these instruments, alongside the other remote sensing and in situ instrument aboard Solar Orbiter will support in our understanding of the characteristics associated with wind emerging from various sources, but are likely not enough. 

To understand the full scope of the source, formation, evolution, acceleration, and heating mechanisms at work in the solar wind we require multi-point in situ observations at different radial and longitudinal separations with complementary remote sensing observations of the wind's coronal origin. While existing fleets of heliophysics spacecraft including Parker, Solar Orbiter, SDO, Hinode, STEREO, and others provide useful in situ and remote sensing observations of varying wind types, the field still struggles to connect remote and in situ observations because inter-spacecraft coordination is limited and comparing in situ measurements and remote observations of the Sun is difficult. Remote and in situ observations can be better connected through multi-point compositional observations alongside the typical plasma parameters measured throughout the heliosphere that offer a robust method for model validation when connecting the solar wind to its source region at the Sun. Remote observations of solar wind source can provide the coronal temperature, density, outflow speeds, heavy ion composition, and magnetic field properties that can be linked to situ measurements of the solar wind, providing critical constraints to formation and acceleration by interchange reconnection or other mechanisms \citep{Rivera-2022wp}. Large scale studies using these sorts of observations will allow us to better probe the source regions of the slow wind and understand the acceleration mechanism(s) at work.

% , more observations allowing for a direct connection between in situ and remote measurements is necessary. Despite all these observatories,

%%%%%%% -------------- ACKNOWLEDGEMENTS -------------- %%%%%%%
\section{Acknowledgements} \label{sec: acknowledgements}

% reviewer!!
The authors would like to thank the reviewer for their comments and suggestions that made the manuscript flow more readable and results more clear and concise.

% FIELDS Team
The FIELDS and SWEAP experiments on the Parker spacecraft was designed and developed under NASA contract NNN06AA01C.

% SWEAP Team
We acknowledge the NASA Parker Solar Probe Mission and the SWEAP team led by J. Kasper for use of data. J. H. acknowledge support of NASA grant 80NSSC23K0737.

% Wind Data
The authors would like to thank the individuals and institutions
who contributed to the Wind mission. Wind data come from
SPDF/CDAWeb. 

% ADAPT Maps
This work utilizes data produced collaboratively between Air Force Research Laboratory (AFRL) \& the National Solar Observatory (NSO). The ADAPT model development is supported by AFRL. The input data utilized by ADAPT is obtained by NSO/NISP (NSO Integrated Synoptic Program). NSO is operated by the Association of Universities for Research in Astronomy (AURA), Inc., under a cooperative agreement with the National Science Foundation (NSF).

% GONG Maps
This work utilizes GONG data obtained by the NSO Integrated Synoptic Program, managed by the National Solar Observatory, which is operated by the Association of Universities for Research in Astronomy (AURA), Inc. under a cooperative agreement with the National Science Foundation and with contribution from the National Oceanic and Atmospheric Administration. The GONG network of instruments is hosted by the Big Bear Solar Observatory, High Altitude Observatory, Learmonth Solar Observatory, Udaipur Solar Observatory, Instituto de Astrofísica de Canarias, and Cerro Tololo Interamerican Observatory.

% SDO Maps
SDO/AIA and SDO/HMI data is courtesy of NASA/SDO and the AIA, EVE, and HMI science teams.

% software
This research used version 4.1.6 of the SunPy open source software package \citep{sunpy}, and made use of HelioPy, a community-developed Python package for space physics \citep{heliopy}. All code to replicate figures can be found at \citet{code-zenodo}. 

\software{
\texttt{Astropy} \citep{astropy:2013, astropy:2018, astropy:2022},
\texttt{heliopy} \citep{heliopy},
\texttt{matplotlib} \citep{mpl},
\texttt{numpy} \citep{numpy},
\texttt{pandas} \citep{pandas},
\texttt{sunkit-magex} \citep{pfss},
\texttt{pySPEDAS} \citep{SPEDAS},
% \texttt{PsiPy} \citep{psipy},
\texttt{scipy} \citep{scipy},
\texttt{spiceypy} \citep{spiceypy},
\texttt{SunPy} \citep{sunpy}
}

%%%%%%% ------------------------------------------------------------------------------------------------------------------------------ %%%%%%%
%%%%%%% -------------------------------------------------------- APPENDIX -------------------------------------------------------- %%%%%%%
%%%%%%% ------------------------------------------------------------------------------------------------------------------------------ %%%%%%%
\appendix
%%% VELOCITY CATEGORIZATION 
\section{Velocity Categorization} \label{appendix: velocity}
Using observations at 1 AU from the Wind 3D Plasma Analyzer (3DP; \citet{Lin-1995aa}), we look at the distribution of velocities over a full solar cycle and for just the time period of interest in this study (January 2020 to December 2022) to look at the fraction of wind at 1 AU that is considered classically \lq{}fast\rq{}. We are interested in doing this to provide a more quantitative cutoff for solar wind speed to differentiate between fast and slow wind at closer heliocentric distances (see Section~\ref{sec: methods-identification}). 

\begin{figure} [ht] %[htb!]
\begin{center}
  \includegraphics[width=\columnwidth]{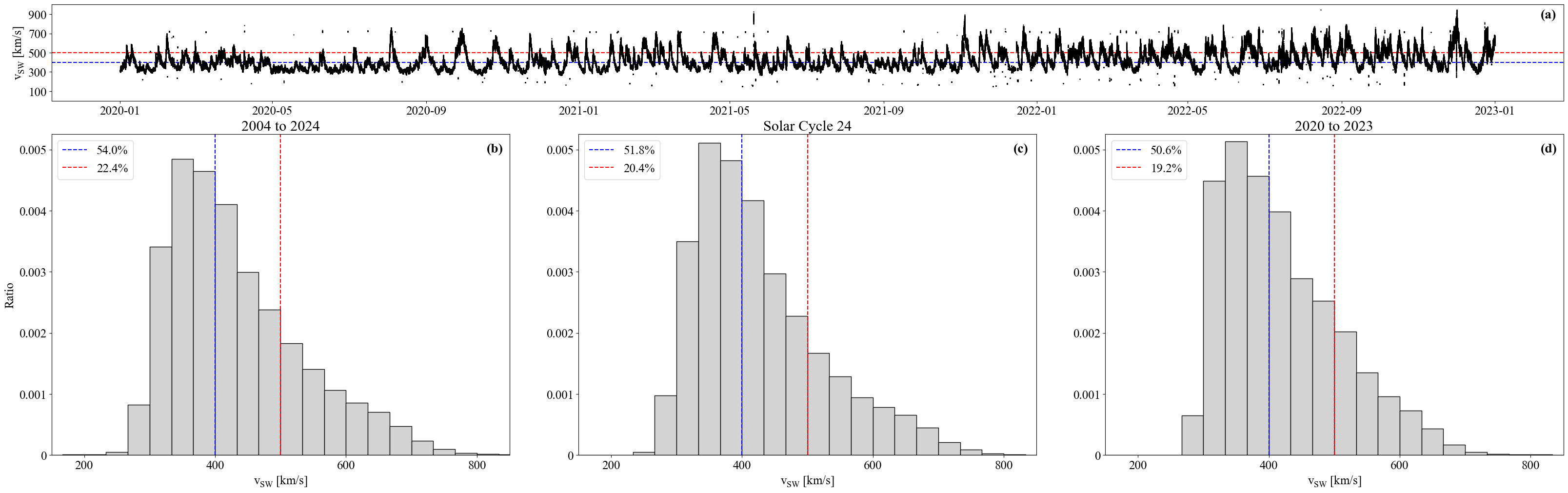}
  \caption{Overview of solar wind velocity observations from the Wind 3DP instrument from 2004 to 2024. 
  \textit{Panel (a):} Time series of solar wind velocity from Wind at 1 AU during the time period of this study.
  \textit{Panel (b):} Distribution of velocity measurements from 2004 to 2024 from Wind.
  \textit{Panel (c):} Distribution of velocity measurements during Solar Cycle 24 (January 2008 to December 2019).
  \textit{Panel (d):} Distribution of velocity measurements from 2020 to 2023, covering the time period in this study. In panels (b) to (d), the dashed red and blue lines indicate the 500~{\kms} and 400~{\kms} speed cutoffs respectively. The legend shows the percentage of observations that fall above that threshold for each period.
  }
  \label{fig: wind_figure}
\end{center}
\end{figure}

Figure~\ref{fig: wind_figure} shows an overview of the distribution function of solar wind velocity measured by Wind. In panel (a), we show a timeseries of the observations spanning from January 2020 to January 2023 (the time period of Parker Encounters 4 to 14). In panels (b), (c), and (d) we look at the distribution functions over different time periods and the percentage of the wind that falls above the 500{\kms} (red) and 400{\kms} (blue) thresholds. Panel (b) shows the last 20 years of observations by Wind, we see that a bit over half of the wind is above 400{\kms} and $\sim 22 \%$ is above 500{\kms}. Looking at just the observations during Solar Cycle 24 (Jan 2008 to December 2019), $\sim 20 \%$ is above 500{\kms} and $\sim 52 \%$ is above 400{\kms}. Looking at just the period of the Parker observations used in this study (panel (d)), we see that about $\sim$20\% of the wind showed speeds above 500 {\kms} and $\sim$50\% above 400 {\kms}. These are the percentage values that we use for our thresholding at described in Section~\ref{sec: methods-identification}. Despite some variance in the solar cycle, these thresholds remain consistent over time giving us confidence in these threshold methods.

%%% STREAM OVERVIEWS
\section{Stream Overviews} \label{appendix: streams}
In this section, we show an overview of the basic plasma parameters and estimated footpoint mapping from PFSS for the main streams studied in this paper. The AIA images used for comparison with the estimated footpoints are chosen based on an estimation of the emergence time of the stream using ballistic mapping of the Parker trajectory with the measured solar wind speed. 

\subsection{Slow {\alfic} Streams}
An overview of the plasma parameters for each of the {\saswnum} SASW streams we more closely examined in this study. Each Figure shows the solar wind speed ($v_R$), absolute cross helicity ($|\sigma_C|$), scaled radial magnetic field ($B_R R^2$), and scaled proton density ($N_P R^2$). The right panel shows a comparison of the estimated footpoints with EUV observations from SDO/AIA. 

\begin{figure} [ht] %[htb!]
\begin{center}
  \includegraphics[width=\columnwidth]{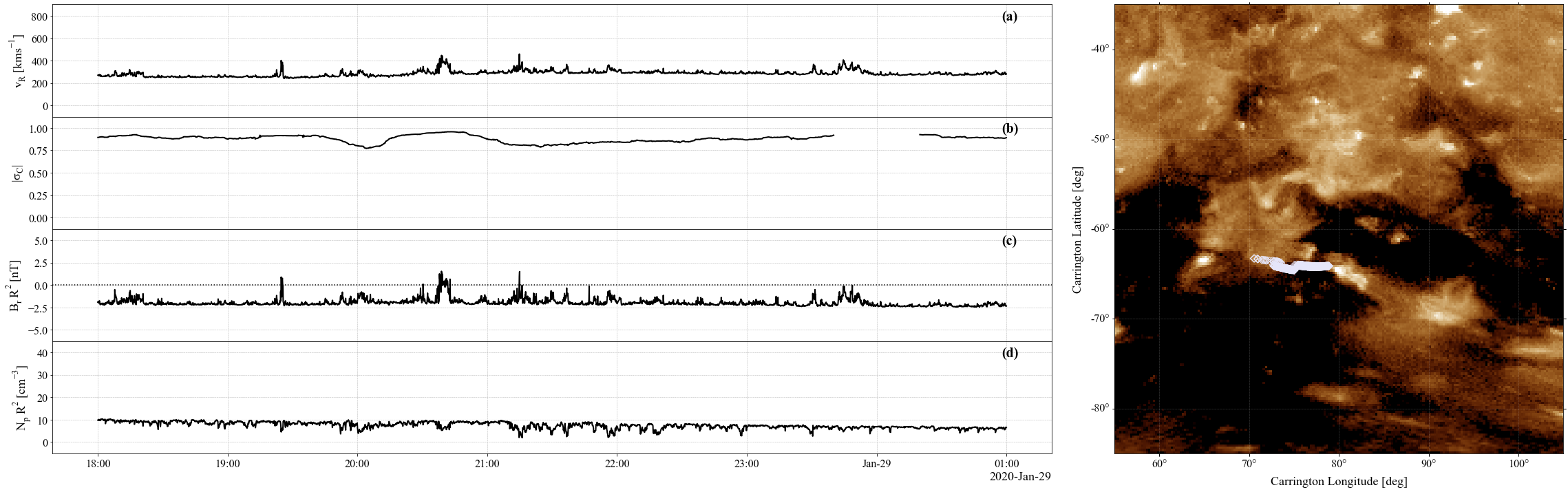}
  \caption{Comparison of in situ characteristics and footpoint estimates for the SASW stream spanning from 2020-01-28 18:00:00 to 2020-01-29 01:00:00 during Encounter Four.
  }
  \label{fig: 01282020_AIA_Data}
\end{center}
\end{figure}

%%%% ENCOUNTER FOUR
Figure~\ref{fig: 01282020_AIA_Data} shows an overview for the SASW stream spanning from 2020-01-28 18:00:00 to 2020-01-29 01:00:00 during Encounter Four. The wind here is seen to emerge from the edge of a coronal at mid-latitudes. The scaled proton density is slightly enhanced over typical 1 AU values (5$\mathrm{cm^{-3}}$).

\begin{figure} [ht] %[htb!]
\begin{center}
  \includegraphics[width=\columnwidth]{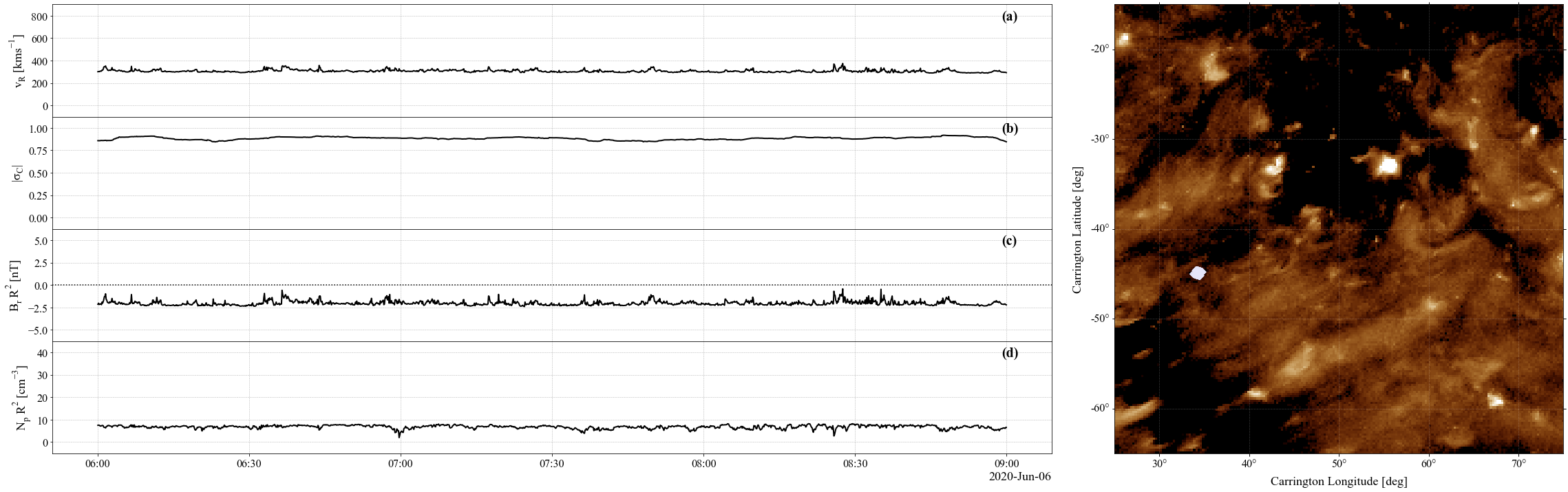}
  \caption{Comparison of in situ characteristics and footpoint estimates for the SASW stream spanning from 2020-06-06 06:00:00 to 2020-06-06 09:00:00 during Encounter Five.
  }
  \label{fig: 06062020_AIA_Data}
\end{center}
\end{figure}

%%%% ENCOUNTER FIVE R1
Figure~\ref{fig: 06062020_AIA_Data} shows an overview for the SASW stream spanning from 2020-06-06 06:00:00 to 2020-06-06 09:00:00 during Encounter Five. The wind here is emerges from a CH edge / small mid-latitude CH. The scaled proton density is slightly enhanced over typical 1 AU values (5$\mathrm{cm^{-3}}$).

\begin{figure} [ht] %[htb!]
\begin{center}
  \includegraphics[width=\columnwidth]{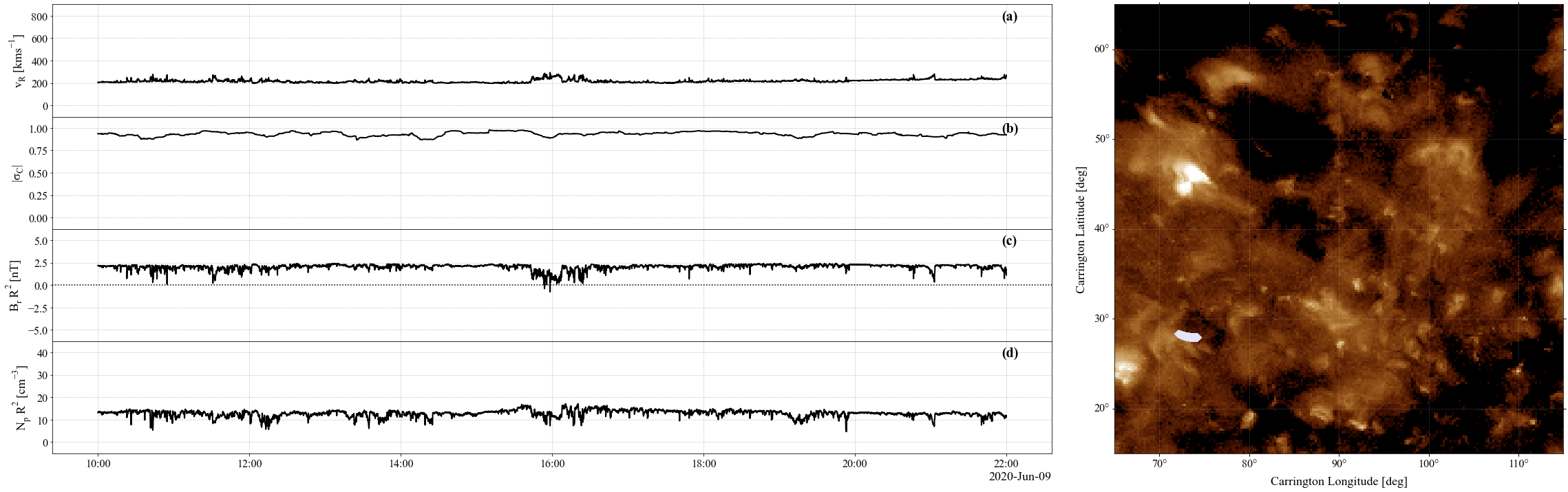}
  \caption{Comparison of in situ characteristics and footpoint estimates for the SASW stream spanning from 2020-06-09 10:00:00 to 2020-06-09 22:00:00 during Encounter Five.
  }
  \label{fig: 06092020_AIA_Data}
\end{center}
\end{figure}

%%%% ENCOUNTER FIVE R2
Figure~\ref{fig: 06092020_AIA_Data} shows an overview for the SASW stream spanning from 2020-06-09 10:00:00 to 2020-06-09 22:00:00 during Encounter Five. The wind here is seen to emerge from a small, mid-latitude CH. The scaled proton density is enhanced over typical 1 AU values (5$\mathrm{cm^{-3}}$).

\begin{figure} [ht] %[htb!]
\begin{center}
  \includegraphics[width=\columnwidth]{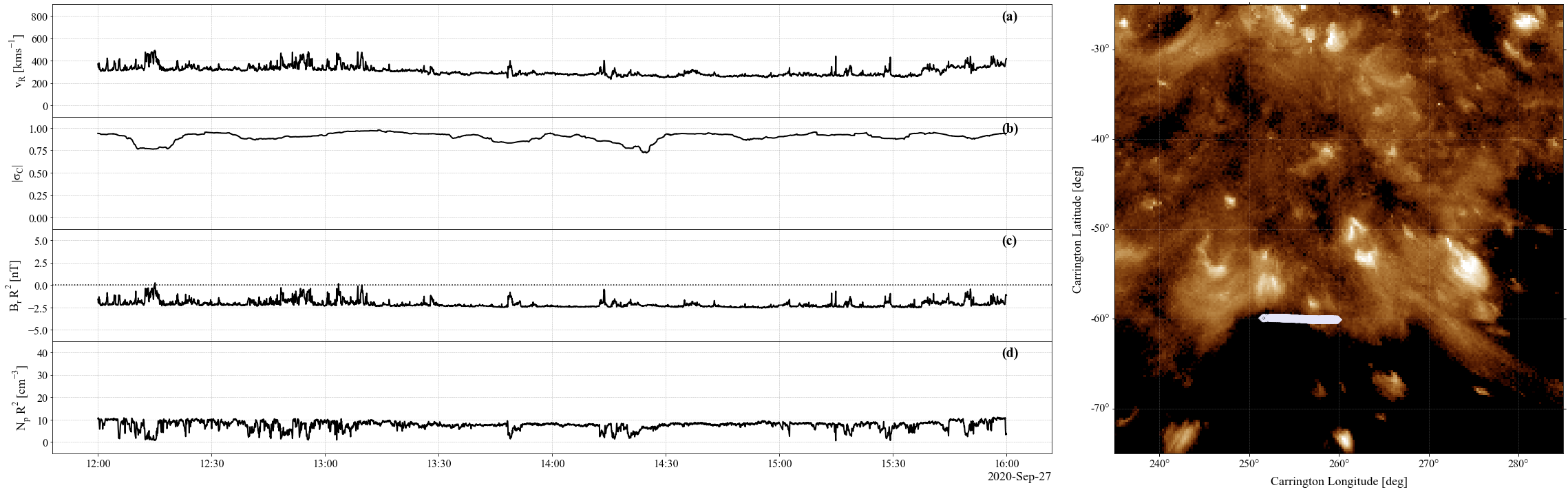}
  \caption{Comparison of in situ characteristics and footpoint estimates for the SASW stream spanning from 2020-09-27 12:00:00 to 2020-09-27 16:00:00 during Encounter Six.
  }
  \label{fig: 09272020_AIA_Data}
\end{center}
\end{figure}

%%%% ENCOUNTER SIX R1
Figure~\ref{fig: 09272020_AIA_Data} shows an overview for the SASW stream spanning from 2020-09-27 12:00:00 to 2020-09-27 16:00:00 during Encounter Six. The wind here is seen to emerge from the edge of a coronal at mid-latitudes. The scaled proton density is slightly enhanced over typical 1 AU values (5$\mathrm{cm^{-3}}$).

%%%% ENCOUNTER SIX R2
% Figure~\ref{fig: 09292020_AIA_Data} shows an overview for the SASW stream spanning from 2020-09-29 02:00:00 to 2020-09-29 05:00:00 during Encounter Six.  The wind here is seen to emerge from the edge of a coronal at mid-latitudes. The scaled proton density is near typical 1 AU values (5$\mathrm{cm^{-3}}$). 

% \begin{figure} [ht] %[htb!]
% \begin{center}
%   \includegraphics[width=\columnwidth]{20200929_E6_AIA_Data.png}
%   \caption{Comparison of in situ characteristics and footpoint estimates for the SASW stream spanning from 2020-09-29 02:00:00 to 2020-09-29 05:00:00 during Encounter Six.
%   }
%   \label{fig: 09292020_AIA_Data}
% \end{center}
% \end{figure}

\begin{figure} [ht] %[htb!]
\begin{center}
  \includegraphics[width=\columnwidth]{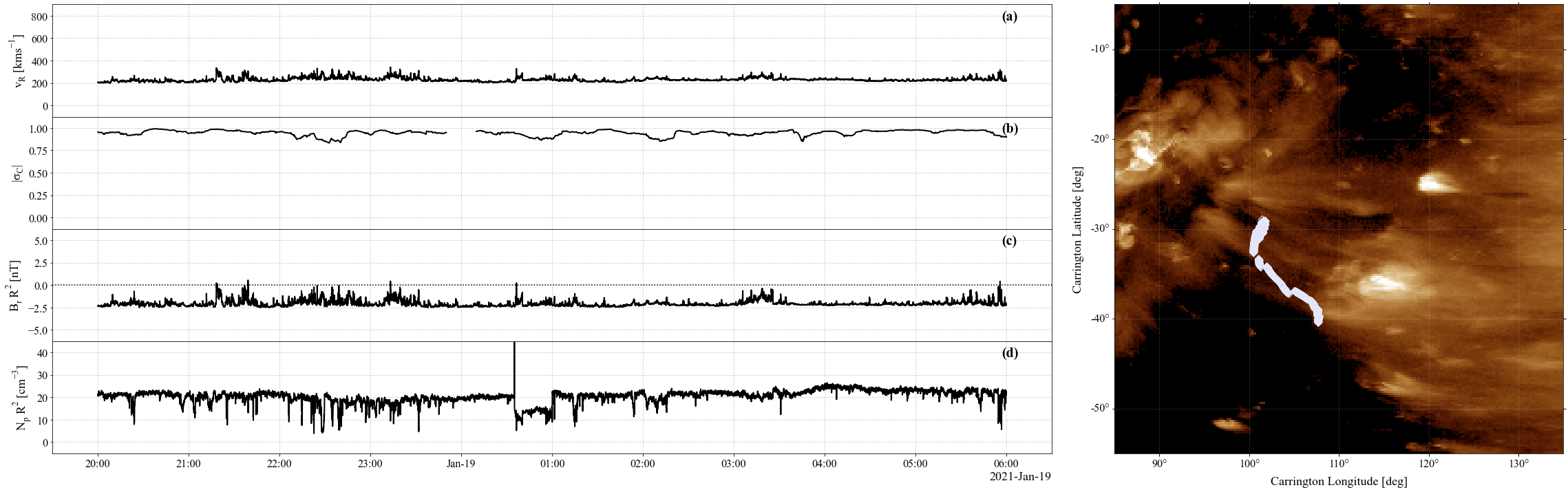}
  \caption{Comparison of in situ characteristics and footpoint estimates for the SASW stream spanning from 2021-01-18 18:00:00 to 2021-01-19 06:00:00 during Encounter Seven.
  }
  \label{fig: 01182021_AIA_Data}
\end{center}
\end{figure}

%%%% ENCOUNTER SEVEN
Figure~\ref{fig: 01182021_AIA_Data} shows an overview for the SASW stream spanning from 2021-01-18 18:00:00 to 2021-01-19 06:00:00 during Encounter Seven. The wind here is seen to emerge from the edge of a coronal at mid-latitudes, with the footpoints outlining the edge of an observed CH. The scaled proton density is enhanced over typical 1 AU values (5$\mathrm{cm^{-3}}$).

\begin{figure} [ht] %[htb!]
\begin{center}
  \includegraphics[width=\columnwidth]{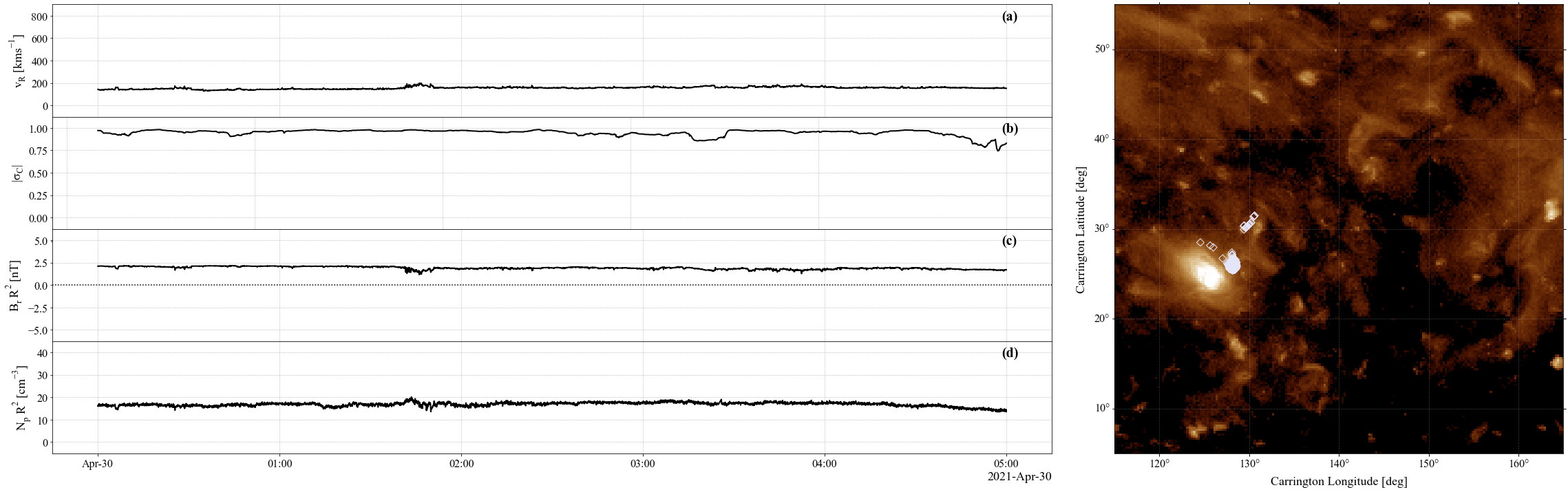}
  \caption{Comparison of in situ characteristics and footpoint estimates for the SASW stream spanning from 2021-04-30 00:00:00 to 2021-04-30 05:00:00 during Encounter Eight.
  }
  \label{fig: 04302021_AIA_Data}
\end{center}
\end{figure}

%%%% ENCOUNTER EIGHT R1
Figure~\ref{fig: 04302021_AIA_Data} shows an overview for the SASW stream spanning from 2021-04-30 00:00:00 to 2021-04-30 05:00:00 during Encounter Eight. This wind stream is very smooth and seems to emerge from the periphery of an active region, with an incredibly incompressible magnetic field, smooth wind speed and enhanced proton density.

\begin{figure} [ht] %[htb!]
\begin{center}
  \includegraphics[width=\columnwidth]{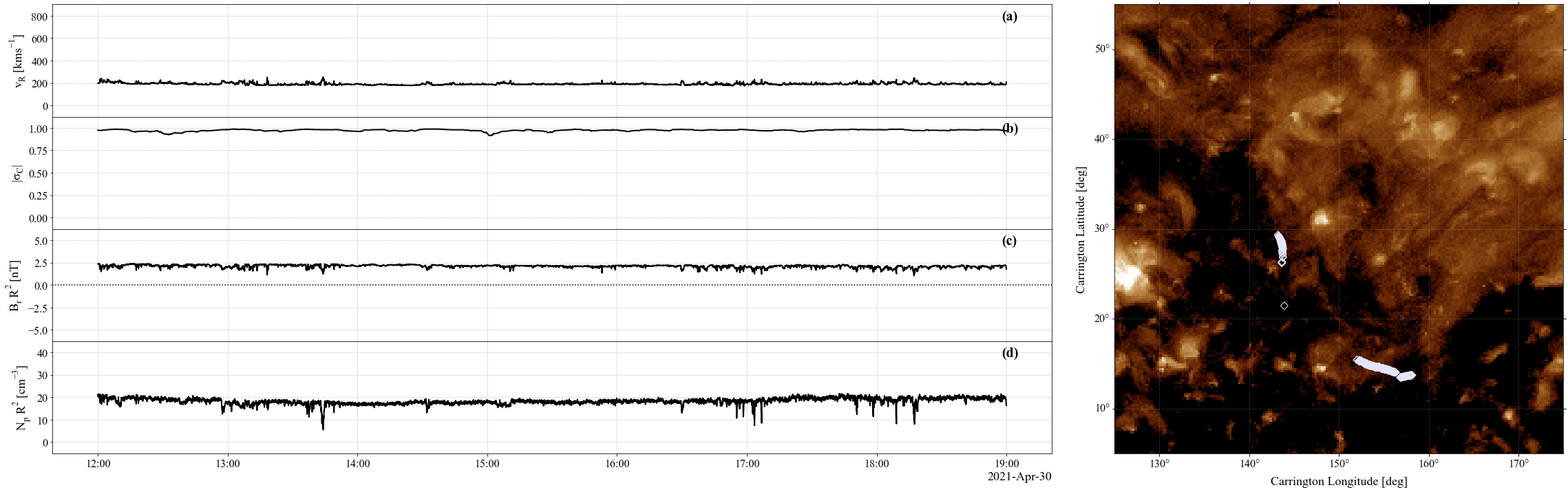}
  \caption{Comparison of in situ characteristics and footpoint estimates for the SASW stream spanning from 2021-04-30 10:00:00 to 2021-05-01 00:00:00 during Encounter Eight.
  }
  \label{fig: 04302021_2_AIA_Data}
\end{center}
\end{figure}

%%%% ENCOUNTER EIGHT R2
Figure~\ref{fig: 04302021_2_AIA_Data} shows an overview for the SASW stream spanning from 2021-04-30 10:00:00 to 2021-05-01 00:00:00 during Encounter Eight. This wind streams emerges from the peripheries of a coronal hole and shows enhanced proton density.

\begin{figure} [ht] %[htb!]
\begin{center}
  \includegraphics[width=\columnwidth]{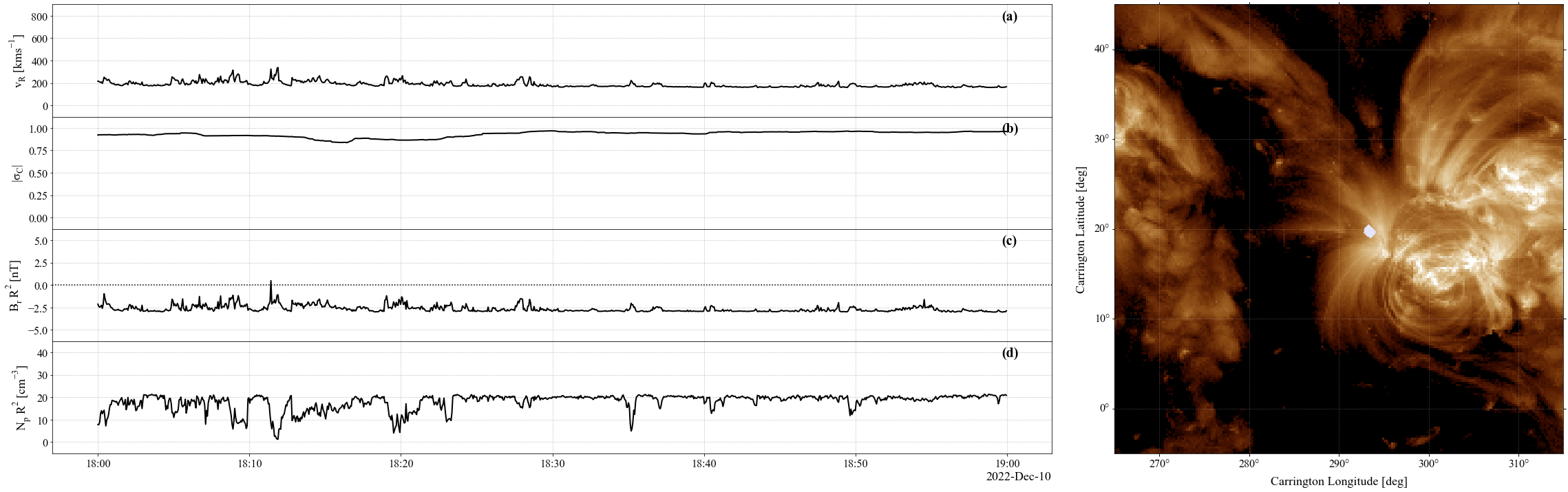}
  \caption{Comparison of in situ characteristics and footpoint estimates for the SASW stream spanning from 2022-12-10 18:00:00 to 2022-12-10 19:00:00 during Encounter Fourteen.
  }
  \label{fig: 12102022_AIA_Data}
\end{center}
\end{figure}

%%%% ENCOUNTER FOURTEEN
Figure~\ref{fig: 12102022_AIA_Data} shows an overview for the SASW stream spanning from 2022-12-10 18:00:00 to 2022-12-10 19:00:00 during Encounter Fourteen. This wind streams shows a large amount of structure, indicative of non-CH wind. It seems to emerge from the edges of a loop like structure at active latitudes and shows structure and enhancement in the proton density.

\begin{figure} [ht] %[htb!]
\begin{center}
  \includegraphics[width=\columnwidth]{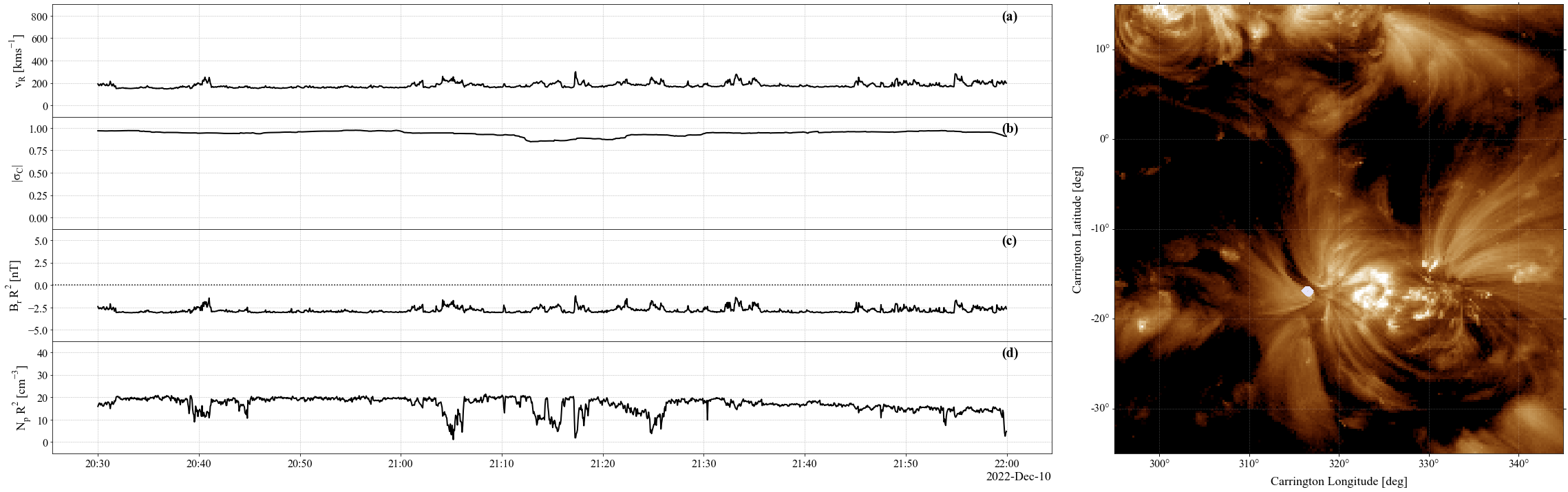}
  \caption{Comparison of in situ characteristics and footpoint estimates for the SASW stream spanning from 2022-12-10 20:30:00 to 2022-12-10 22:00:00 during Encounter Fourteen.
  }
  \label{fig: 12102022_2030_22_AIA_Data}
\end{center}
\end{figure}

%%%% ENCOUNTER FOURTEEN
Figure~\ref{fig: 12102022_2030_22_AIA_Data} for the SASW stream spanning from 2022-12-10 20:30:00 to 2022-12-10 22:00:00 during Encounter Fourteen. This wind streams shows a large amount of structure, indicative of non-CH wind. It seems to emerge from the edges of a loop like structure at active latitudes and shows structure and enhancement in the proton density.

\begin{figure} [ht] %[htb!]
\begin{center}
  \includegraphics[width=\columnwidth]{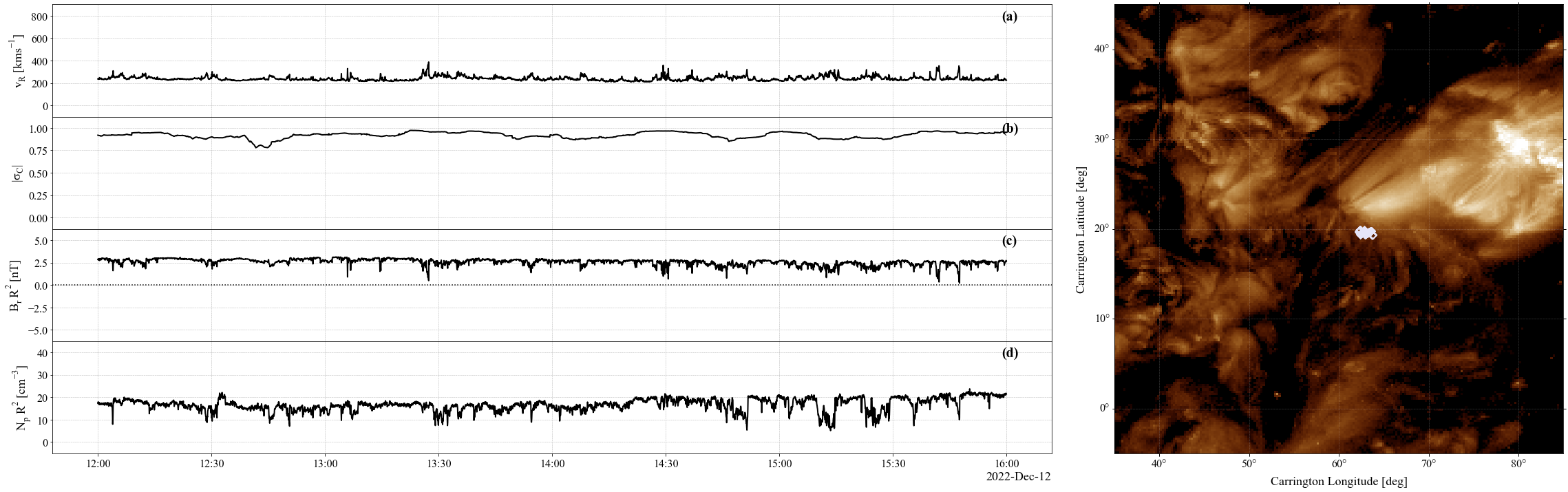}
  \caption{Comparison of in situ characteristics and footpoint estimates for the SASW stream spanning from 2022-12-12 12:00:00 to 2022-12-12 16:00:00 during Encounter Fourteen.
  }
  \label{fig: 12122022_AIA_Data}
\end{center}
\end{figure}

%%%% ENCOUNTER FOURTEEN
Figure~\ref{fig: 12122022_AIA_Data} shows an overview for the SASW stream spanning from 2022-12-12 12:00:00 to 2022-12-12 16:00:00 during Encounter Fourteen. This wind streams shows enhanced proton density and looks to emerge from the periphery of a coronal hole near an active region structure. There is structure in the velocity, magnetic field, and observed proton density.

\subsection{Fast Wind Streams}
An overview of the plasma parameters and footpoint mapping for the {\fswnum} FSW streams modeled for this study. These five streams all emerge from within or near the boundaries of coronal hole structures.

\begin{figure} [ht] %[htb!]
\begin{center}
  \includegraphics[width=\columnwidth]{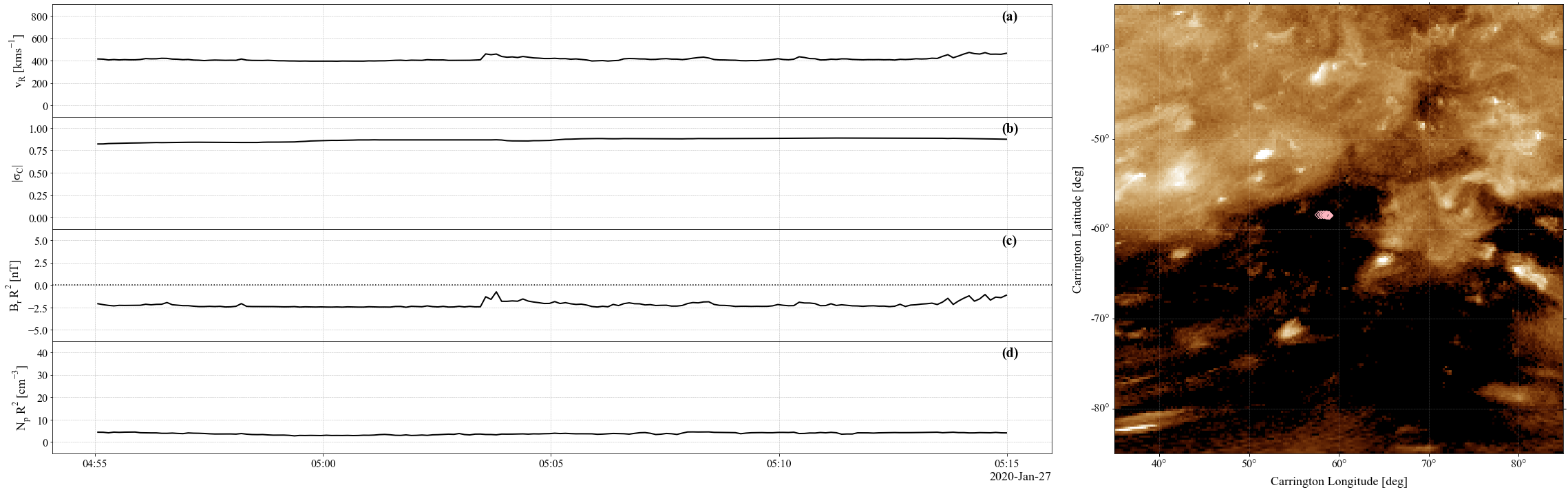}
  \caption{Comparison of in situ characteristics and footpoint estimates for the FSW stream spanning from 2020-01-27 04:55:00 to 2020-01-27 05:15:00 during Encounter Four.
  }
  \label{fig: 2020-01-27_FSW_AIA_Data}
\end{center}
\end{figure}

%%%% ENCOUNTER FOUR
Figure~\ref{fig: 2020-01-27_FSW_AIA_Data} shows an overview for the SASW stream spanning from 2020-01-27 04:55:00 to 2020-01-27 05:15:00 during Encounter Four.

\begin{figure} [ht] %[htb!]
\begin{center}
  \includegraphics[width=\columnwidth]{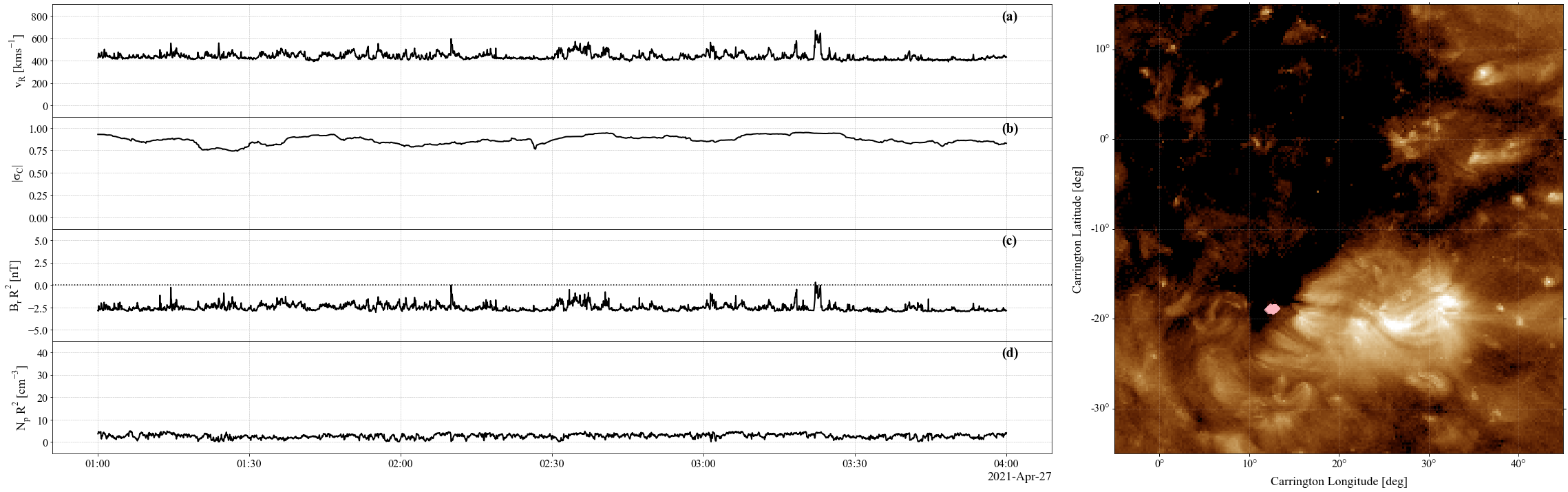}
  \caption{Comparison of in situ characteristics and footpoint estimates for the FSW stream spanning from 2021-04-27 01:00:00 to 2021-04-27 04:00:00 during Encounter Eight.
  }
  \label{fig: 2021-04-27_FSW_AIA_Data}
\end{center}
\end{figure}

%%%% ENCOUNTER EIGHT
Figure~\ref{fig: 2021-04-27_FSW_AIA_Data} shows an overview for the SASW stream spanning from 2021-04-27 01:00:00 to 2021-04-27 04:00:00 during Encounter Eight.

\begin{figure} [ht] %[htb!]
\begin{center}
  \includegraphics[width=\columnwidth]{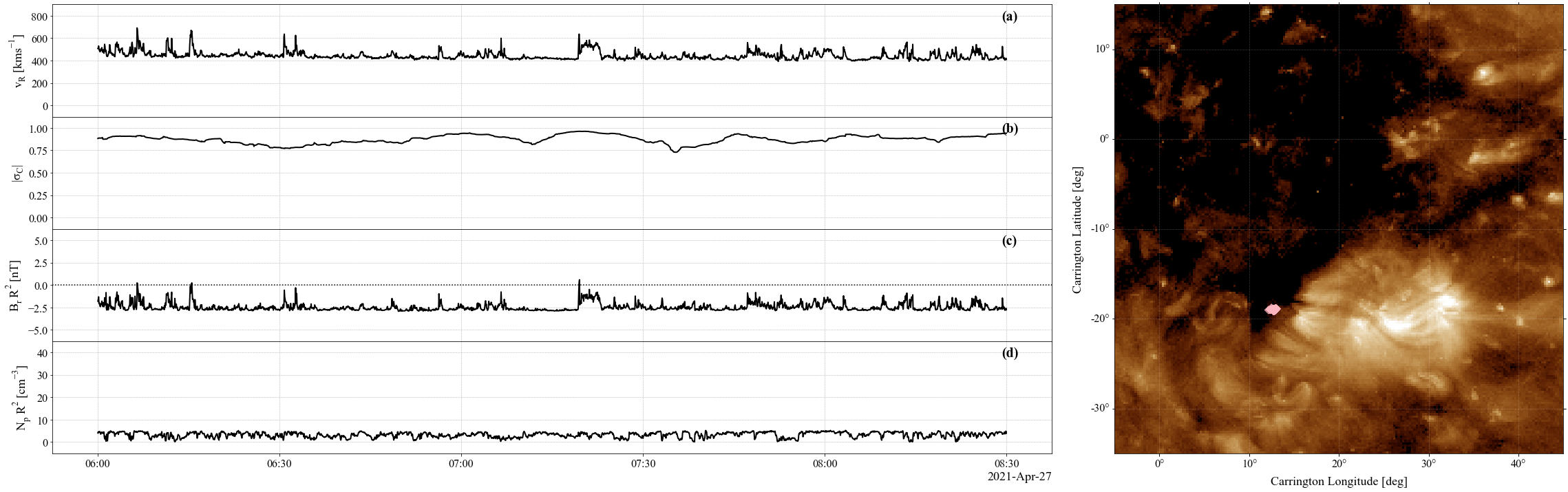}
  \caption{Comparison of in situ characteristics and footpoint estimates for the FSW stream spanning from 2021-04-27 06:00:00 to 2021-04-27 08:30:00  during Encounter Eight.
  }
  \label{fig: 2021-04-27_2_FSW_AIA_Data}
\end{center}
\end{figure}

%%%% ENCOUNTER EIGHT
Figure~\ref{fig: 2021-04-27_2_FSW_AIA_Data} shows an overview for the SASW stream spanning from 2021-04-27 06:00:00 to 2021-04-27 08:30:00  during Encounter Eight.

\begin{figure} [ht] %[htb!]
\begin{center}
  \includegraphics[width=\columnwidth]{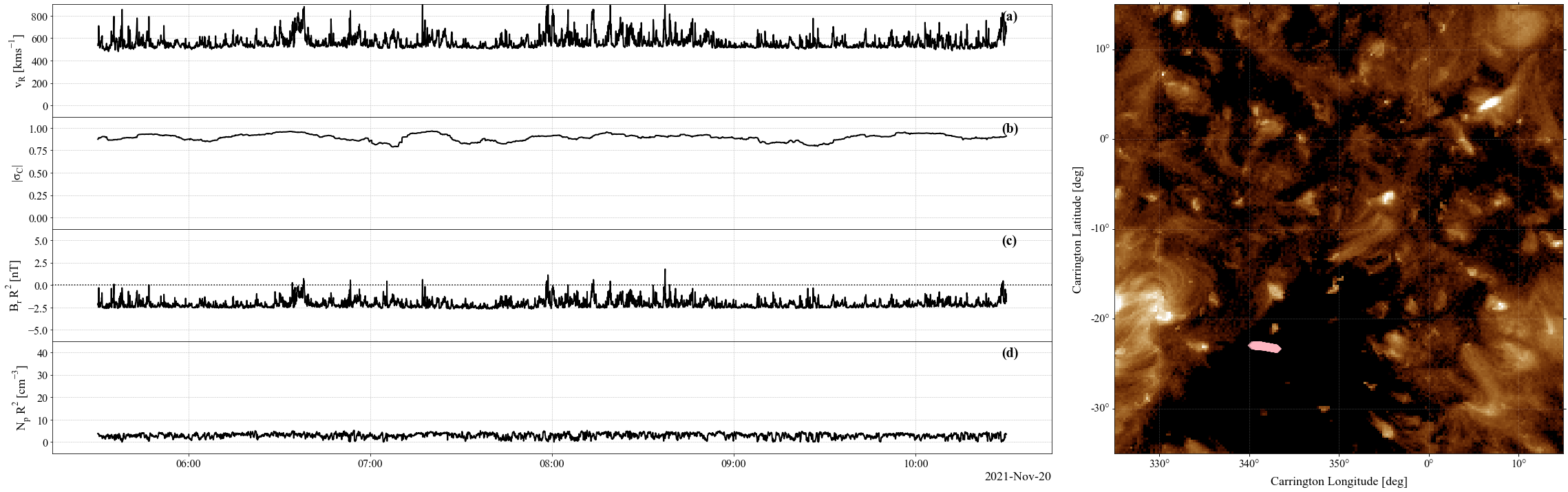}
  \caption{Comparison of in situ characteristics and footpoint estimates for the FSW stream spanning from 2021-11-20 05:30:00 to 2021-11-20 10:30:00 during Encounter Ten.
  }
  \label{fig: 2021-11-20_FSW_AIA_Data}
\end{center}
\end{figure}

%%%% ENCOUNTER TEN
Figure~\ref{fig: 2021-11-20_FSW_AIA_Data} shows an overview for the SASW stream spanning from 2021-11-20 05:30:00 to 2021-11-20 10:30:00 during Encounter Ten.

\begin{figure} [ht] %[htb!]
\begin{center}
  \includegraphics[width=\columnwidth]{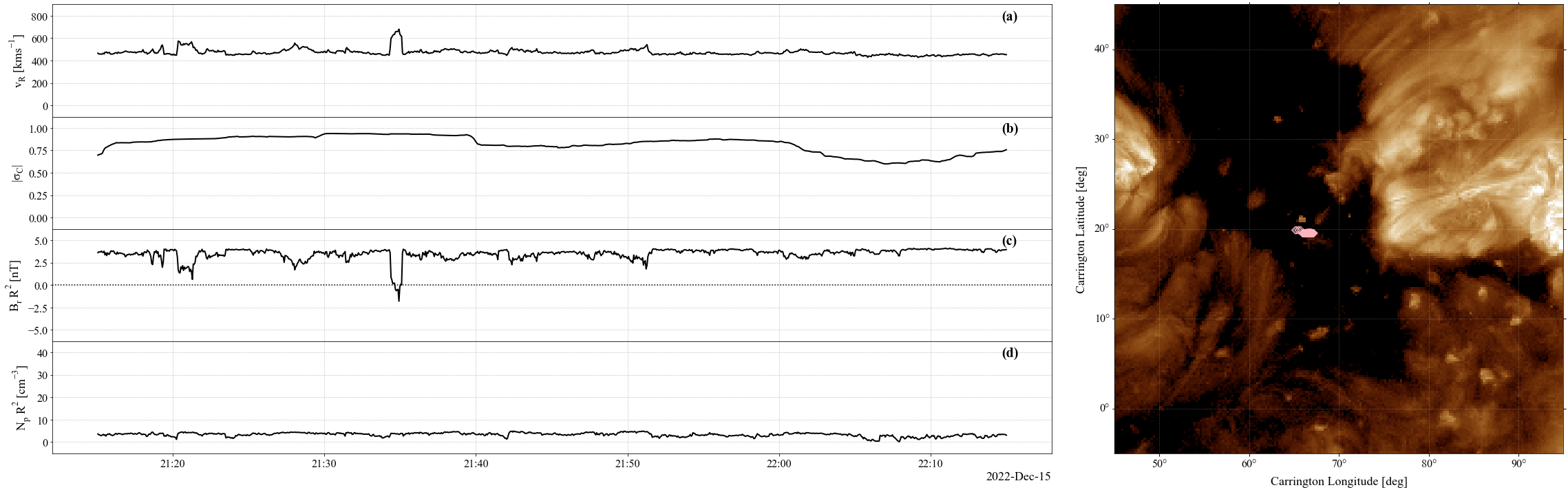}
  \caption{Comparison of in situ characteristics and footpoint estimates for the FSW stream spanning from 2022-12-15 21:15:00 to 2022-12-15 22:15:00 during Encounter Fourteen.
  }
  \label{fig: 2022-12-14_FSW_AIA_Data}
\end{center}
\end{figure}

%%%% ENCOUNTER FOURTEEN
Figure~\ref{fig: 2022-12-14_FSW_AIA_Data} shows an overview for the SASW stream spanning from 2022-12-15 21:15:00 to 2022-12-15 22:15:00 during Encounter Fourteen. This stream comes from the center of a coronal hole and has scaled proton densities $\mathrm{\sim 5 cm^{-3}}$.

%%%%%%% -------------- Modeling Validation -------------- %%%%%%%
\section{Model Validation} \label{appendix: validation}

%%% overview 
This study uses the Potential Field Source Surface (PFSS) model alongside ballistic propagation methods to estimate the footpoints of plasma streams of interest. The PFSS model has two inputs that can lead to variance in the overall results: the choice of input magnetogram and source surface height. Similar to methods previously outlined in \citet{Badman-2020, Badman-2022, Ervin-2024CH, Ervin-2024SA} we look at the effect of these two parameters to understand the impact on our conclusions. 

%%% ballistic propagation
The ballistic propagation results depend on the in situ $v_R$ measurement and methodology, and the results of this projection can impact the estimated footpoints. We probe how variance in the input velocity effect our footpoint estimations \citep{Ervin-2024CH, Ervin-2024SA}. We vary the inputted in situ $v_R$ used to calculate the source surface trajectory via ballistic propagation by bootstrapping noise on at a level of 4\% -- the estimated systematic uncertainty in the SPAN-I velocity moment (3\%) plus some extra noise to account for random sources of error \citep{Livi-2022}. From these new source surface trajectories,  we calculate the estimated footpoint from PFSS with the \lq{}best\rq{} magnetogram used for identifying the source region of each SASW and FSW stream. There are a variety of source of noise and uncertainty in the SPAN-I velocity measurements and this methodology aims to show how noise and uncertainty in the velocity measurement affect the resulting source surface trajectory and footpoints. 

\begin{figure} [ht] %[htb!]
\begin{center}
  \includegraphics[width=\columnwidth]{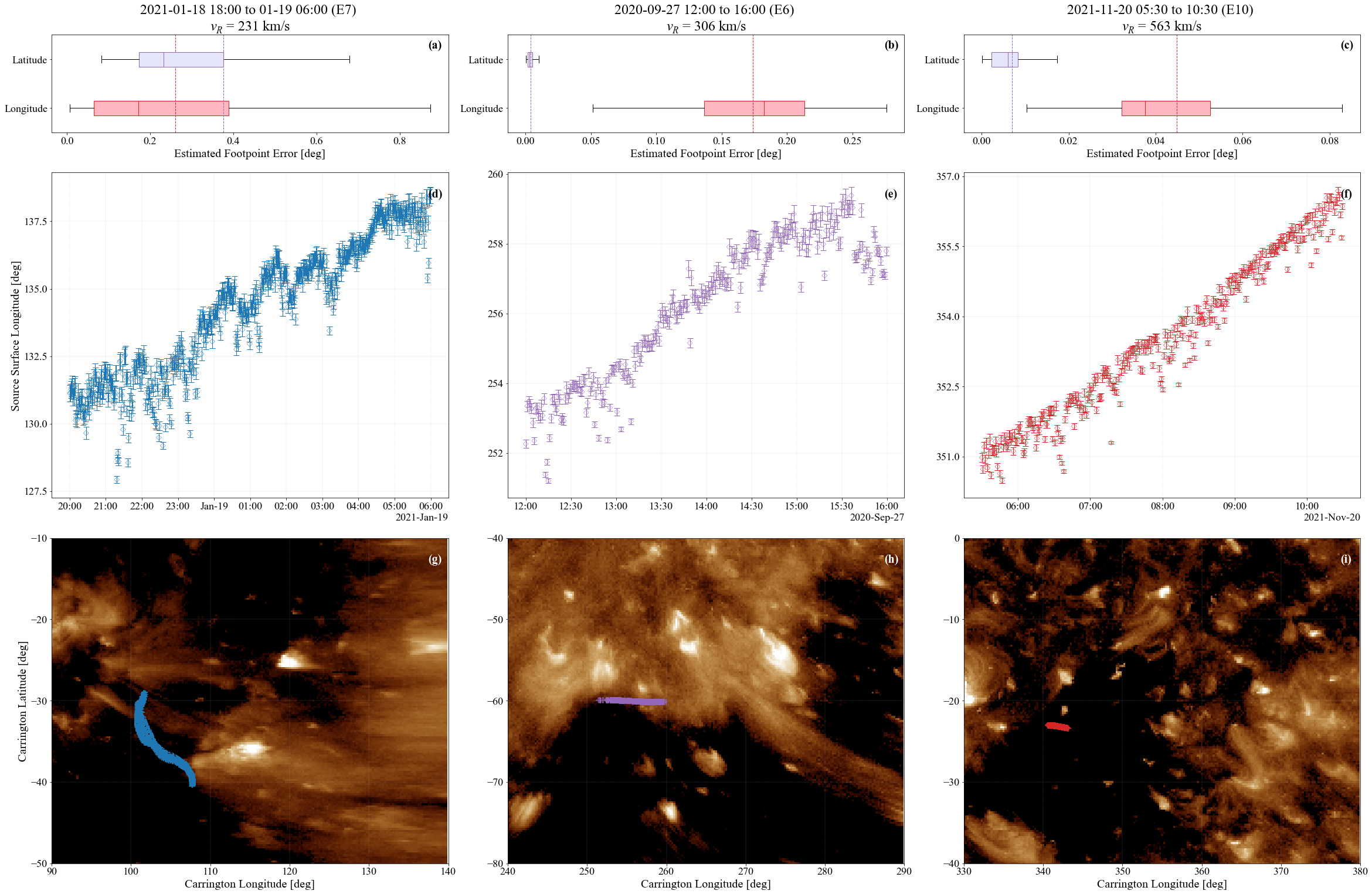}
  \caption{Overview of error on estimated PFSS footpoints due to noise in the velocity measurement used for ballistic propagation. We use three streams of different speeds to show how the effect varies based on {\vsw}. The top row (panels (a), (b), (c)) show the overall error on the latitude and longitude of the footpoints. The middle row (panels (d), (e), (f)) show the error on the source surface longitude as a function of time. The bottom row (panels (g) to (i)) shows the error on the footpoints compared with the associated AIA image. 
  }
  \label{fig: ballistic_noise}
\end{center}
\end{figure}

In Figure~\ref{fig: ballistic_noise}, we show an overview of the variance in the source surface trajectory when adding noise of 4\% to the in situ $v_R$. We choose three streams to represent different characteristics speeds: fast, slow, and slower, and show how the slower wind speeds are more affected by the inputted velocity to the ballistic propagation. This is one contributing factor to the difficulty in determining the sources of the slow wind in comparison to the source of faster streams. Figure~\ref{fig: ballistic_noise} shows how the uncertainty in the velocity translates to \lq{}error\rq{} on the estimated footpoints determined from PFSS. In the top row we show the distribution of error on our estimated footpoints finding it to be less than 1{\degree} for these periods. In panels (d) to (f) we show how the error in ballistic propagation impacts the resulting source surface trajectory and finally the footpoint estimation (panels (g) to (i)).

We see that while the exact coordinate position in terms of longitude/latitude has some slight variance (see panels (a) to (c) and (g) to (i)), the overall source region remains unaffected. There is an obvious correlation between the speed of the wind stream and the effects on ballistic propagation, notably that there is more error in slow wind streams than in fast, a contributor to the issues surrounding modeling the true source of slow wind. \citet{Dakeyo-2024} found similar results noting that the estimated error was $\leq 5^\circ$ and that the error decreased with heliocentric distance and had an inverse correlation with wind speed.

%%%% FIGURE --- ballistic noise

\begin{figure} [ht] %[htb!]
\begin{center}
  \includegraphics[width=\columnwidth]{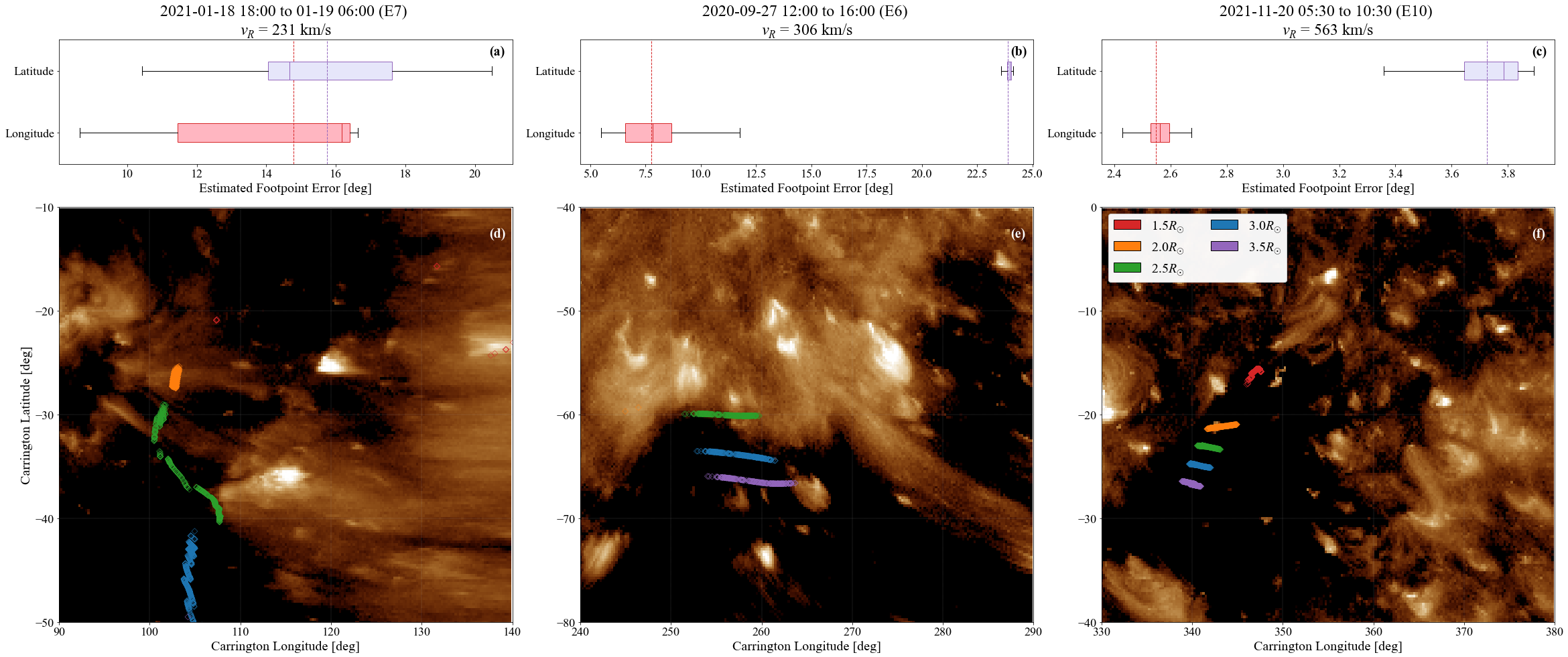}
  \caption{Comparison of the error on the estimated footpoints due to changing the source surface height for streams of varying speeds. Panels (a) to (c) show the overall error on the footpoints as a box plot. Panels (d) to (f) show the footpoints associated with different source surface heights on the associated AIA image. 
  }
  \label{fig: source_surface}
\end{center}
\end{figure}

\begin{figure} [ht] %[htb!]
\begin{center}
  \includegraphics[width=\columnwidth]{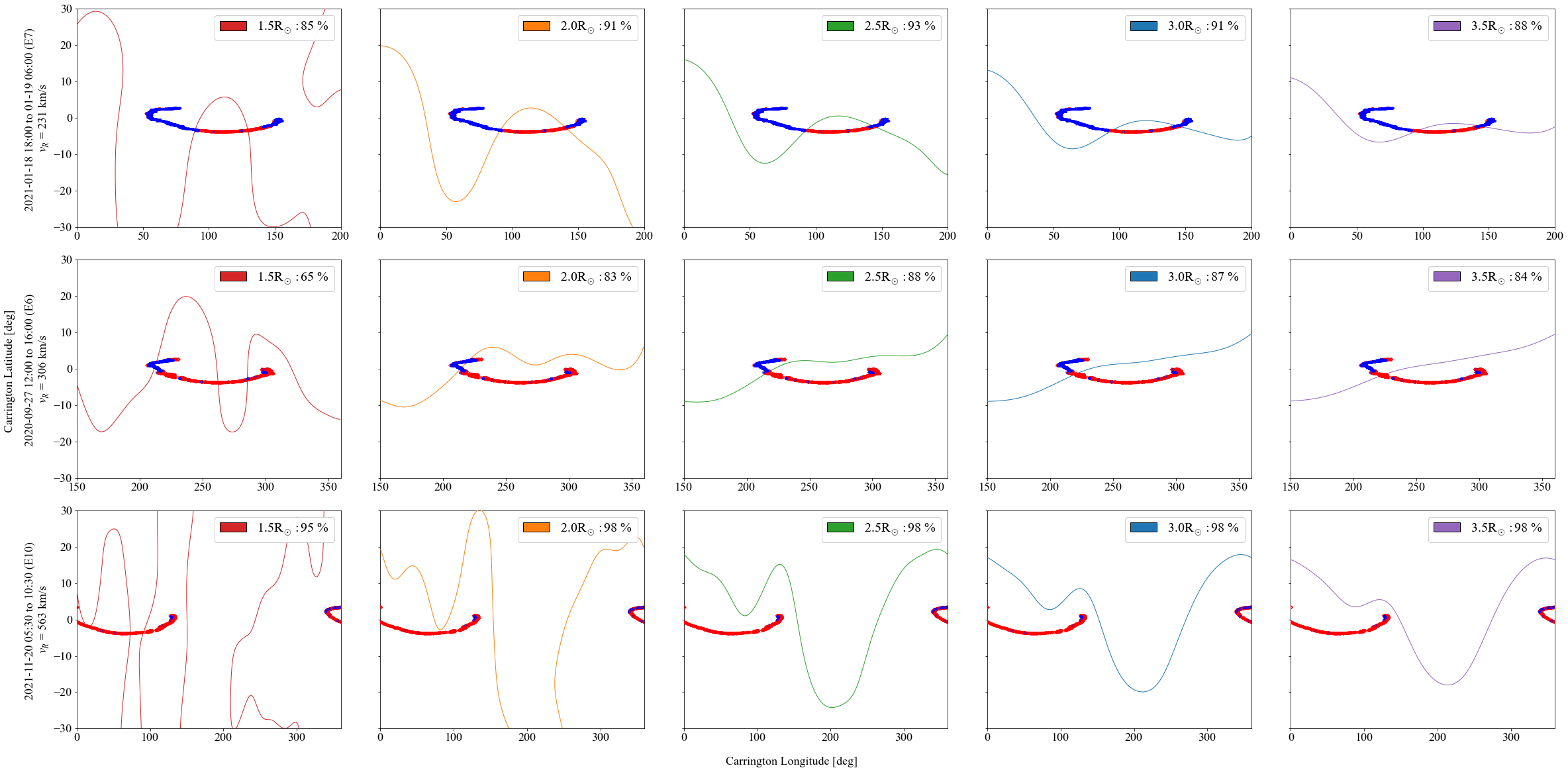}
  \caption{Validation of our choice of source surface height ({\Rss}) for the three different streams we have discussed in this section. The rows corresponding to the streams from E7, E6, and E10 from top to bottom respectively. The trajectory of Parker at the source surface is shown in blue (negative polarity) and red (positive polarity) based on the measurements from Parker/FIELDS. The heliospheric current sheet from the PFSS model is compared with the trajectory and we show the percentage of correct polarity the model predicted in the legend.
  }
  \label{fig: polarity_val}
\end{center}
\end{figure}

%%% source surface height
In the PFSS model, the chosen source surface height can impact the outputted footpoints as this height ({\Rss}) determines that point at which the field is forced to be purely radial. For each of the streams of interest, we vary the height of the source surface between 2.0~{\Rsun} and 3.5~{\Rsun} to understand how this impacts the resulting footpoints. We see in Figure~\ref{fig: source_surface} that changing the source surface height leads to large variance in the location of the footpoints, we see from Figure~\ref{fig: polarity_val} that the optimal {\Rss} is 2.5~{\Rsun}. Varying {\Rss} impacts the exact position of the footpoint (especially in latitude), however it does not impact the resulting source region. It does have a larger effect on the resulting footpoints in comparison with the effect of the error of the in situ $v_R$ measurement. We use a source surface height of 2.5~{\Rsun} for our modeling as this produces the best result in terms of matching the modeled polarity with the measured magnetic field polarity from FIELDS. Similar to the error in the ballistic propagation, we see that this most impacts slower wind streams.

%%%% FIGURE --- source surface height

%%%% FIGURE --- polarity validation

%%% input magnetogram
The last user input to PFSS, the choice of input magnetogram, also impacts the resulting footpoint. We choose a \lq{}best\rq{} magnetogram by comparing the outputted polarity from the PFSS model to the observed polarity by Parker and optimizing such that the neutral line crossings match. To quantify the impact on the estimated footpoints by this choice, we use magnetograms from $\pm$2 days around our choice of magnetograms with a {\Rss} of 2.5~{\Rsun} and compare results. We find that this produces an error on the estimated footpoint of $\sim$2{\degree} and thus is an important consideration when validating the PFSS model.

%%%% FIGURE --- input magnetogram
\begin{figure} [ht] %[htb!]
\begin{center}
  \includegraphics[width=\columnwidth]{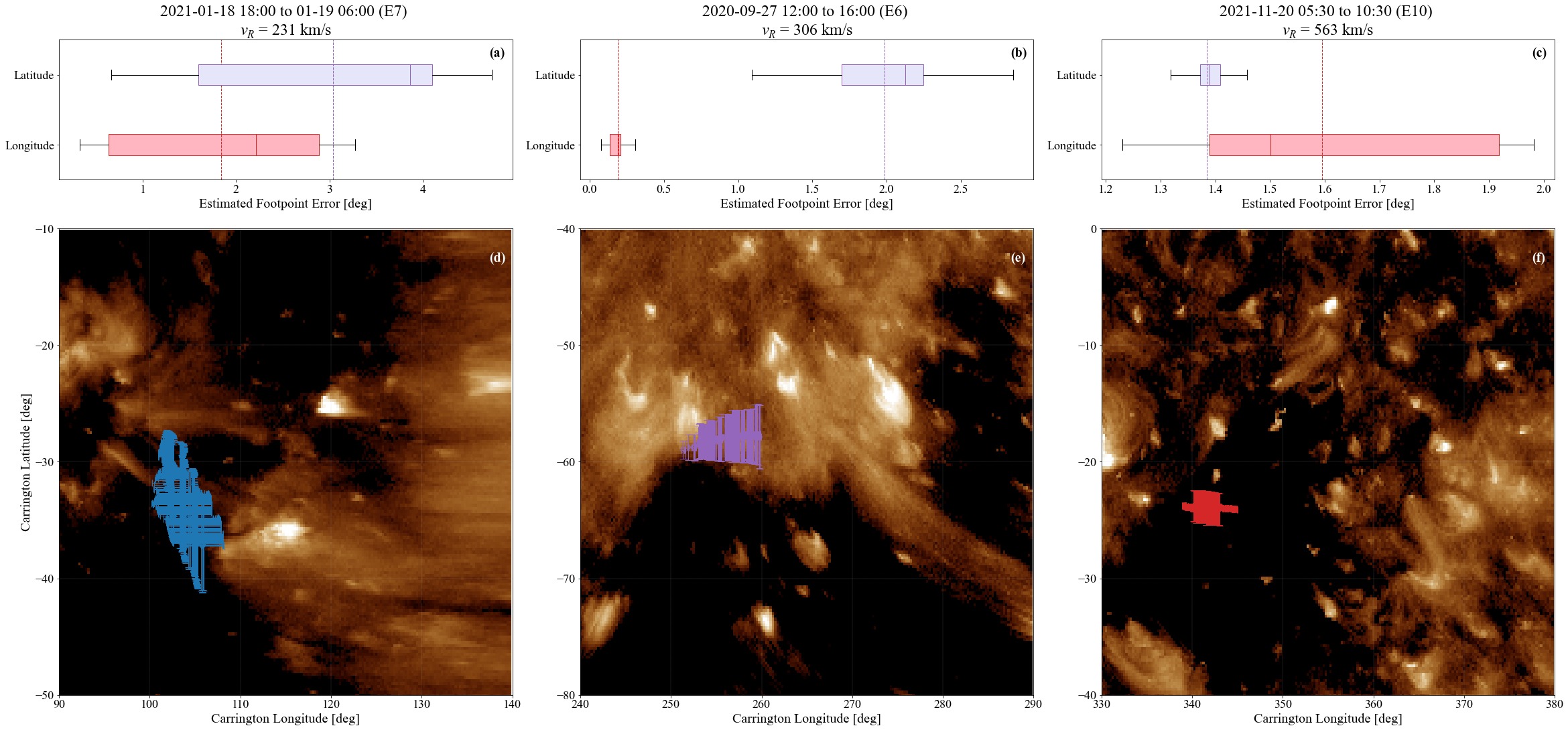}
  \caption{Comparison of the effect of the choice of input magnetogram for the lower PFSS boundary condition on the estimated footpoints for streams of varied speeds. Panels (a) to (c) show the overall error on the estimated footpoint for each of the wind streams of interest. Panels (d) to (f) show the footpoints and their error on the associated AIA images.
  }
  \label{fig: input_mag}
\end{center}
\end{figure}

Through this analysis of the ballistic propagation and PFSS processes required to trace from the spacecraft down to estimated photospheric footpoints, we see that there are a variety of error sources which can impact the results. Most notably, the source surface height has the largest impact on the resulting footpoints. Much work has been done looking at this height, comparing it with other models, optimizing to reproduce the observed open flux, and considering a non-spherical source surface to better reproduce observations. We see that the impact of the errors in the ballistic propagation and PFSS are considerably less impactful for the fast wind, further contributing to the difficulties in determining the source of the slow wind. While we can validate our models with magnetic field observations, multi point observations would provide additional measurements for further constraints and confidence in our results. In this work, we look at statistical distributions of parameters and while overlaps in distributions could be attributable to the error sources described above, it would not the systematic shift in source region properties seen in Figure~\ref{fig: footpoint_expansion}. This indicates that the observed differences are real and due to variance source regions of the wind.

%%%%%%% ------------------------------------------------------------------------------------------------------------------------------ %%%%%%%
%%%%%%% -------------------------------------------------------- REFERENCES -------------------------------------------------------- %%%%%%%
%%%%%%% ------------------------------------------------------------------------------------------------------------------------------ %%%%%%%
\bibliography{ms}{}
\bibliographystyle{aasjournal}
\end{document}